\documentclass[11pt,a4paper]{article}
\usepackage{jheppub}
\usepackage{natbib}
\usepackage{graphicx}
\usepackage{cases}
\usepackage[dvipsnames]{xcolor}

\usepackage{mathtools}
\usepackage{hyperref}
\usepackage{float}
\usepackage{comment}
\usepackage{multirow}

\def\nat{Nature}

\def\mnras{MNRAS}

\def\aap{A\&A}

\definecolor{Blu}{rgb}{0.,0.,1.}

\author[a,b]{Sunghyun Kang,}
\author[a,b]{Arpan Kar,}
\author[a,b]{Stefano Scopel,}
\affiliation[a]{Center for Quantum Spacetime, Sogang University, 35 Baekbeom-ro, Mapo-gu, Seoul, 121-742, South Korea}
\affiliation[b]{Department of Physics, Sogang University, 35 Baekbeom-ro, Mapo-gu, Seoul, 121-742, South Korea}
\emailAdd{francis735@naver.com}
\emailAdd{arpankarphys@gmail.com}
\emailAdd{scopel@sogang.ac.kr}

\title{Halo--independent bounds on Inelastic Dark Matter}

\abstract{We discuss halo--independent constraints on the Inelastic Dark Matter (IDM) scenario, in which a Weakly Interaction Massive Particle (WIMP) state $\chi$ with mass $m_\chi$ interacts with nuclear targets by upscattering to a heavier state $\chi^{\prime}$ with mass $m_\chi+\delta$.
In order to do so we adopt the single--stream method, that exploits the complementarity of Direct Detection (DD) and Capture in the Sun to extend the experimental sensitivity to the full range of incoming WIMP speeds.
We show that a non--vanishing mass splitting $\delta$ modifies such range, and that for particular combinations of $m_\chi$ and $\delta$ the complementarity between the two detection techniques required by the method is lost. In such cases a specific choice of the WIMP speed distribution in our Galaxy is required to obtain a constraint on the WIMP--nucleus cross section or coupling. Specifically, assuming for the escape velocity in our Galaxy $u_{\rm esc}$ the reference value $u_{\rm esc}^{ref}$ = 560 km/s a halo--independent bound is possible when $\delta\lesssim$ 510 keV for a Spin--Independent interaction and when $\delta\lesssim$ 245 keV for a Spin--Dependent interaction (with the Spin--Independent value slightly reduced to $\delta\lesssim$ 490 keV when $u_{\rm esc}>u_{\rm esc}^{ref}$). In the low WIMP mass regime the bound from capture in the Sun is always more constraining than that for DD and is sufficient alone to provide a halo--independent constraint, while for large WIMP masses the halo--independent bound is given by a combination of capture in the Sun and DD. In this latter case the reduction in sensitivity due to the mass splitting $\delta$ is more pronounced for DD than for capture, and capture plays a more important role to determine the halo--independent bound compared to the elastic case. In particular we find that, for $u_{\rm esc}$ = $u_{\rm esc}^{ref}$, unless the mass of the target used in DD is larger than about four times that of the target driving capture in the Sun, DD does not play any role in the determination of the maximal value of $\delta$ for which a halo--independent bound is possible. We also discuss the issue of thermalization of IDM within the Sun and show that its impact on our results is mild.} 

\begin{document}
\hspace*{87.0mm}{CQUeST-2023-0723}\\
\maketitle

\section{Introduction}
\label{sec:introduction}

Weakly Interacting Massive Particles (WIMPs) are the most natural candidates to provide the Cold Dark Matter that is supposed to have triggered galaxy formation and is believed to provide about 25\% of the density of the Universe under 
an invisible form~\cite{Planck:2018vyg}, only detected so far through its gravitational effects.
The main process that can be used to search for WIMPs is their scattering process off nuclear targets, that enters at the same time in Direct Detection (DD) experiments, that search for the recoil energy of nuclei in solid--state, 
liquid and gaseous detectors in underground laboratories shielded 
against cosmic rays~\cite{DD_Goodman1984, DD_LEWIN1996, JUNGMAN1996, DD_Schumann2019, Snowmass_Leane2022}, 
or in experiments searching for neutrinos produced by WIMP annihilation inside celestial bodies (Earth, Sun), 
where the WIMPs are accumulated after being captured through 
the same WIMP--nucleus scattering process 
that enters DD~\cite{cap_nu_sun_PhysRevLett1985, cap_nu_sun_HAGELIN1986, dm_cap_nu_SREDNICKI1987, Jungman:1994jr, idm_sun_Catena_2018}. 

An important piece of input in the calculation of expected signals is the WIMP speed distribution $f(u)$ in the reference frame of the Solar system, that determines the WIMP incoming flux\footnote{Neglecting the relative velocity between the Earth and the Sun both present direct detection experiments and signals from WIMP capture in the Sun are sensitive to the speed distribution $f(u) \equiv \int d\Omega f(\vec{u}) u^2$.}. 
In particular, for the $f(u)$ both early analytical estimations~\cite{violent_relaxation} and more recent numerical models of Galaxy formation~\cite{VDF_Lacroix2020, VDF_Lopes2020} are compatible to a Maxwellian in the galactic halo rest frame~\cite{SHM_1986, SHM_1988}, at least for speeds that are not far larger than a speed dispersion estimated to be of the order of $\simeq$ 300 km/s from the measurement of the galactic rotation curve, assuming hydrodynamic equilibrium between the pressure of the WIMP gas and the gravitational pull toward the center. This simple scenario, that predicts a flat rotation curve in agreement with observation, is also indicated as the Standard Halo Model (SHM). However, although the SHM provides a useful zero--order approximation to describe the WIMP speed distribution, numerical simulations of Galaxy formation can only shed light on statistical average properties of galactic halos, whilst our lack of information about the specific merger history of the Milky Way prevents us to rule out the possibility that the $f(u)$ has sizeable non--thermal components. Indeed, the growing number of observed dwarf galaxies hosted by the Milky Way suggests that our halo is 
not perfectly thermalized~\cite{Gaia_2018Nature, Gaia_2018MNRAS, Gaia_Myeong_2018, Gaia_Koppelman_2019, Gaia_Necib_2019, Gaia_Necib_2020, Gaia_OHare_2020}, 
and the more so should be expected in the high--speed tail of the $f(u)$ to which, for instance, DD signals are particularly sensitive for light WIMP masses~\cite{DEAP2020, DD_Gaia_Bozorgnia2019}.
Based on the above considerations several attempts have been made to develop halo--independent approaches with the goal to remove the dependence of the experimental bounds on the choice of a specific speed distribution $f(u)$~\cite{halo_independent_2010, halo_independent_Fox_2010, halo_uncertainty_Frandsen2011, astrophysics_independent_Herrero-Garcia2012, halo_independent_DelNobile_2013, Bozorgnia:2013hsa, halo_independent_Fox2014, halo_independent_Feldstein2014, halo_independent_Scopel_inelastic_2014, halo_independent_Feldstein2014_2, halo_independent_Bozorgnia2014, halo_independent_Anderson2015, Halo-independent_Ferrer2015, halo_independent_Kahlhoefer, Gondolo_Scopel_2017, halo_independent_Catena_Ibarra_2018, velocity_uncertainty_Ibarra2018, velocity_independent_2019}.

The techniques listed above have been mainly developed in the context of direct detection with the goal to assess in a halo--independent way the compatibility of an experimental excess, such as the DAMA modulation effect~\cite{dama_libra_phase2}, with the constraints from other detectors~\cite{halo_independent_DelNobile_2013, halo_independent_Catena_Ibarra_2018}.  On the other hand, in absence of a clear excess an alternative strategy is to work out the most conservative bounds from null searches compatible with the only constraint:

\begin{equation}
    \int_{u=0}^{\infty} f(u) du = 1,
    \label{eq:f_normalization}
\end{equation}

\noindent but allowing for any possible speed profile of the distribution.

In the case of this latter approach WIMP direct searches run into a crucial limitation: all DD experiments are characterized by a recoil energy threshold $E_R^{\rm th}$ that for a given WIMP mass $m_\chi$ converts into a speed threshold $u^{\rm DD}_{\rm th}$ below which the sensitivity to the WIMP flux vanishes. As a consequence no conservative bounds can be established from existing DD experiments because the latter cannot probe the full range of WIMP speeds. In particular any functional form of the $f(u)$ for which $\int_{u=0}^{u^{\rm DD}_{\rm th}} f(u) du = 1$, and, consequently, $\int_{u=u^{\rm DD}_{\rm th}}^{\infty} f(u) du = 0$, corresponds to a vanishing expected signal in all existing DD experiments.

A possible solution to this problem is provided by combining the constraints from DD with those from the expected neutrino signal from WIMPs captured in the Sun~\cite{NT_DD_Kavanagh2014, NT_DD_Blennow2015}. 
Indeed, while capture in the Sun is suppressed at high WIMP incoming speeds, it is favoured for low (even vanishing) ones, because in the latter case it is easier for a slow WIMP to be scattered below the escape speed in order to remain gravitationally trapped in the celestial body. Such complementarity between DD and capture in the Sun allows to probe experimentally the $f(u)$ distribution in the full range of WIMP incoming velocities, and
was exploited in Ref.~\cite{Halo-independent_Ferrer2015} to develop a particularly straightforward method that allows to obtain conservative constraints~\cite{halo_independent_sogang_2023} that are independent of the $f(u)$ and only require the assumption~(\ref{eq:f_normalization}). For convenience, in the following we will refer to such procedure as the ``single stream method", and to the ensuing constraints as ``single--stream halo--independent" bounds.

Both DD and capture are substantially modified in the scenario of Inelastic Dark Matter~\cite{inelastic_Tucker-Smith:2001} (IDM). In this class of models a Dark Matter (DM) particle $\chi$ of mass $m_\chi$ interacts with atomic nuclei exclusively by upscattering to a second heavier state $\chi^{\prime}$ with mass $m_{\chi^{\prime}}$ = $m_\chi$ + $\delta$. A peculiar feature of IDM is that there is a minimal WIMP incoming
speed in the target frame matching the kinematic threshold for inelastic upscattering events and given by:

\begin{equation}
    v_{T*}=\sqrt{\frac{2\delta}{\mu_{\chi T}}},
    \label{eq:vstar}
\end{equation}

\noindent with $\mu_{\chi T}$ the reduced mass of the WIMP and the nuclear target $T$. The WIMP speed in the halo of our Galaxy is bounded by the escape velocity $u_{\rm esc}\simeq$ 560 km/s~\cite{vesc_Smith2006, vesc_Piffl2013} and for both DD and capture in the Sun or the Earth the targets move in the Galactic rest frame with a relative speed $v_0\simeq v_\odot$ with $v_\odot$ the rotational curve at the solar system position, for which in the following we will assume $v_\odot$ = 220 km/s~\cite{SHM_maxwell_Green2011}.  As a consequence the WIMP--scattering process in DD vanishes if $v_{T*}> u_{\rm max} = u_{\rm esc} + v_0\simeq$ 800 km/s, because it becomes kinematically not accessible, the incoming WIMP speed being too slow to overcome the inelasticity threshold. Such kinematic bound implies that the largest values of $\delta$ that DD can probe are reached by the heaviest targets (xenon, iodine, tungsten) employed in existing experiments and cannot exceed $\delta\simeq$ 200 keV~\cite{iDM_DD_Nagata2014}. The sensitivity to the $\delta$ parameter can be improved beyond that for DD by looking for processes where the WIMP scatters off nuclei at higher speeds.  For instance, the gravitational potential accelerates the WIMP particles before they scatter off the nuclei of a celestial body, so that their speed can be larger than $v_{\rm max}$. Indeed, it can reach 1600 km/s in the center of the Sun, implying that the values of $\delta$ that can be probed in IDM by capture can reach $\simeq$ 600 keV~\cite{idm_sun_Catena_2018}. As a consequence, for these scenarios the solar capture bounds are stronger than those from DD. Even stronger bounds can in principle be obtained from WIMP capture in celestial bodies with a gravitational potential stronger than that of the Sun, such as white dwarves~\cite{wd_sogang_2023} or neutron stars~\cite{IDM_neutron_stars_2018, IDM_neutron_stars_2022, Nguyen_2022}.

The main goal of the present paper is to assess the validity and applicability of the halo--independent single--stream method to the case of Inelastic Dark Matter\footnote{In our analysis we will only assume $\delta>0$. It is worth pointing out that when $\delta<0$ (exothermic DM~\cite{exothermic_dm_graham_2010}) or in the scenario of fermionic DM absorption~\cite{DM_absorbtion_2020} the expected rate can be almost halo--independent in the first place, when the energy released from the rest mass of the absorbed DM state outweighs its kinetic energy.}. In particular, it is clear from the discussion above that a non--vanishing mass splitting $\delta$ modifies the range of incoming velocities $u$ both DD and Capture in the Sun are sensitive to. As we will show in our analysis, this will imply that for particular combinations of $m_\chi$ and $\delta$ the complementarity of DD and Capture in the Sun required by the single--stream method is lost, in the sense that the combination of the two detection techniques is not sensitive anymore to the full range of $u\le u_{\rm max}$. In such cases a specific choice of the $f(u)$ will be required to obtain a constraint. 

Another aspect that we wish to discuss is how inelasticity affects the complementarity between DD and Capture. Indeed, while $\delta>$0 increases the velocity threshold of the scattering process and reduces the available phase space for both DD and capture, eventually making both kinematically inaccessible, this is expected to affect more DD than capture. In fact, for capture the incoming WIMP is faster thanks to the gravitational acceleration in the celestial body. Moreover, the endothermicity of the inelastic process reduces the kinetic energy of the final states for the same WIMP incoming speed, hindering DD, where the only detected signal is the recoil energy of the nucleus $E_R$, but affecting much less capture, where instead $E_R$ has no lower threshold. 

On the other hand some issues arise when capture of IDM in the Sun is considered. In particular the threshold speed of Eq.~(\ref{eq:vstar}) for inelastic upscattering events can easily exceed the velocity dispersion that the WIMPs need to have in order to be in thermal equilibrium with the solar plasma, an assumption that is commonly made to calculate their annihilation rate. This issue was addressed with a numerical simulation in the specific case of a Maxwellian distribution~\cite{Blennow_2018}. In our analysis we will calculate expected signals assuming for definiteness thermal equilibrium between the WIMPs and the solar plasma, and show that the effect of giving up this assumption is relatively mild. 

As pointed out in~\cite{Halo-independent_Ferrer2015, halo_independent_sogang_2023} the single--stream method can only be applied when the expected rate is proportional to a single coupling or cross section. In the following, we will discuss the two popular cases of a
spin--independent (SI) WIMP--nucleus cross section, with a scattering amplitude proportional to the atomic mass number of the target, or  a spin--dependent (SD) interaction with a scattering amplitude proportional to the coupling between the spins of the WIMP and of the nucleus. Both SI and SD arise in popular extensions of the Standard model, such as Supersymmetry~\cite{JUNGMAN1996}, and 
represent the operators at zero--th order in the WIMP--nucleon relative velocity of the most general WIMP--nucleon effective Hamiltonian compatible to Galilean invariance for a particle of spin 1/2~\cite{nreft_haxton1, nreft_haxton2, all_spins_sogang_2021}.

The plan of the paper is the following. In Section~\ref{sec:idm_scattering} we summarize the main features of the  WIMP--nucleus scattering process for both capture in the Sun (Section~\ref{sec:capture}) and DD (Section~\ref{sec:dd}); in Section~\ref{sec:combined_sensitivity_idm} we discuss in detail the combined kinematics of DD and capture in the Sun in the case of IDM, showing for which combinations of $m_\chi$ and $\delta$ the full range of WIMP incoming speeds $u$ is probed and for which it is not. In Section~\ref{sec:single_stream} we outline the procedure to quantitatively determine the halo--independent single--stream bound, and in Section~\ref{sec:analysis} we apply it to calculate the halo--independent single--stream exclusion plots of the WIMP--proton and WIMP--neutron SI and SD couplings combining the DD experimental null searches of LUX--ZEPLIN(LZ)~\cite{LZ_2022}, XENON1T~\cite{xenon_2018}, PICO--60 ($C_3F_8$)~\cite{pico60_2019} and PICO--60 ($CF_3I$)~\cite{pico60_2015} with those from the Super--Kamiokande~\cite{SuperK_2015} and IceCube~\cite{IceCube:2016} neutrino telescopes, and provide the corresponding ranges of $m_\chi$ and $\delta$ for which they are possible. Such results represent the main output of our analysis. Our conclusions are contained in Section~\ref{sec:conclusions}. Moreover, in Appendix~\ref{app:experiments} we provide the details about the implementation of the experimental bounds, while in Appendix~\ref{app:thermalization} we discuss the impact on our results of assuming thermalization of IDM within the Sun. Finally, since for definiteness in Sections~\ref{sec:combined_sensitivity_idm} and \ref{sec:analysis} the reference value $u_{\rm max}^{ref}$ = 780 km/s is adopted for the maximal WIMP speed $u_{\rm max}$, in Appendix~\ref{app:umax_dep} we discuss what happens when $u_{\rm max}>u_{\rm max}^{ref}$.

\section{WIMP--nucleus scattering}
\label{sec:idm_scattering}

The WIMP--nucleon interaction can be parameterized in terms of an effective non--relativistic Hamiltonian that 
parameterizes the most general WIMP--nucleus interaction invariant by Galilean transformations~\cite{nreft_haxton1,nreft_haxton2}. In the present work we will only consider the standard spin--independent and spin--dependent interactions which, following the notation of \cite{nreft_haxton1,nreft_haxton2}, are driven by the following operators: 
\begin{equation}
\mathcal{O}_1 = 1_\chi 1_{\cal N} \,\, ({\rm SI}) \,\,, 
\hspace{8mm} 
\mathcal{O}_4 = \vec{S}_\chi \cdot \vec{S}_{\cal N} \,\, ({\rm SD}) \,,
\label{eq:O1_O4}
\end{equation}
where $1_\chi$ and $1_{\cal N}$ are identity operators and $\vec{S}_\chi$ and $\vec{S}_{\cal N}$ are the spin operators for the WIMP and for the nucleon. In the following we will indicate with $c_1$ and $c_4$ the SI and SD couplings, respectively, expressed in units of GeV$^{-2}$.

In the general case the WIMP--nucleus scattering amplitude can be written as~\cite{nreft_haxton1,nreft_haxton2}:
\begin{equation}
\frac{1}{2 j_{\chi}+1} \frac{1}{2 j_{T}+1}|\mathcal{M}_T|^2 = 
\frac{4\pi}{2 j_{T}+1} \sum_{\tau=0,1}\sum_{\tau^{\prime}=0,1} \sum_{k} 
R^{\tau\tau^{\prime}}_k
\left [(c^{\tau})^2, (v^\perp)^2, q^2\right ] W_{T k}^{\tau\tau^{\prime}}(q),
\label{eq:squared_amplitude}
\end{equation}
where $j_{\chi}$ and $j_{T}$ are the WIMP and the target nucleus spins respectively, 
$(v^\perp)^2=v^2-v_{\rm min}^2$ with $v$ the incoming WIMP speed and $v_{\rm min}$ the minimal value of $v$ required to impart the nuclear recoil events, and $q=|\vec{q}|$ is the magnitude of the transferred momentum. 
The Wilson coefficients in the isospin base $c^{\tau}$ (with $\tau = 0, 1$)  can be 
converted into the couplings of the WIMP with protons and neutrons through $c^p=(c^0+c^1)/2$ and $c^n=(c^0-c^1$)/2. 
As already pointed out in our analysis we will only consider the two cases $c^{\tau}$ = $c^{\tau}_1$ (SI interaction) and 
$c^{\tau}$ = $c^{\tau}_4$ (SD interaction), in the notation of Ref.~\cite{nreft_haxton1,nreft_haxton2}. 
The functions $R^{\tau\tau^{\prime}}_k$ and $W^{\tau\tau^{\prime}}_{T k}$ are the WIMP response function and the nuclear response function (or nuclear form factor), respectively, and can be found in Refs.~\cite{nreft_haxton1,nreft_haxton2,Catena_nuclear_form_factors} (in particular for the SI and the SD interactions the WIMP response functions $R_k$ do not depend explicitly on $(v^\perp)^2$ and $q^2$).
The index $k$ indicates different effective nuclear operators and 
for the standard SI and SD interactions are $k = M$ and $k = \Sigma^{\prime\prime}, \Sigma^{\prime}$, respectively.

Finally, the differential cross section for the WIMP--nucleus scattering process
in terms of the recoil energy ($E={q^2}/{2 m_T}$) is given by:
\begin{equation}
\frac{d\sigma_T}{d E}=\frac{2 m_T}{4\pi v^2}\left [ \frac{1}{2 j_{\chi}+1} \frac{1}{2 j_{T}+1}|\mathcal{M}_T|^2 \right ],
\label{eq:dsigma_de}
\end{equation}
\noindent with $m_T$ the mass of the nuclear target.

\subsection{Capture in the Sun}
\label{sec:capture}

WIMPs can be gravitationally captured in the Sun through their inelastic scattering ($\chi + T \rightarrow \chi^\prime + T$) off the nuclear targets in its interior. 
The captured WIMPs can annihilate in pairs to produce several final states among which only a flux of energetic neutrinos can escape the solar plasma and be detected on Earth. The bounds on the neutrino flux from the Sun constrain the WIMP annihilation rate, given by:

\begin{equation}
    \Gamma_{\odot} = \frac{C_{\odot}}{2} \tanh^2 (t_\odot/\tau_\odot)
    \label{eq:gamma_rate}
\end{equation}

\noindent where $t_\odot$ is the age of the Sun, while the equilibration time $\tau_\odot$ is given by:

\begin{equation}
\tau_\odot= (C_\odot C_A)^{-\frac{1}{2}},     
\label{eq:t_eq}
\end{equation}

\noindent with~\cite{griest_seckel_1986}:

\begin{eqnarray}
    &&C_A=\frac{\langle \sigma v \rangle 4\pi \int_{0}^{R_\odot} r^2 n_\chi^2(r)\,dr}{\left [  4\pi \int_{0}^{R_\odot} r^2 n_\chi(r)\,dr\right ]^2}\label{eq:ca}.
    \label{eq:CA}
\end{eqnarray}
    
In the equation above $n_\chi$ is the density of WIMPs within the Sun, $\langle\sigma v\rangle$ is the WIMP annihilation cross section times velocity, for which in the following we will adopt the standard value $\langle\sigma v\rangle$ = $3\times 10^{-26}$ cm$^3$ s$^{-1}$, which, assuming s--wave annihilation, corresponds to the observed relic density for a thermal WIMP. Finally $C_\odot$ is the capture rate, which in the optically thin limit is given by~\cite{idm_sun_Catena_2018, idm_sun_Shu_2010, idm_sun_Menon_2009, idm_sun_Nussinov_2009}:
\begin{equation}
C_\odot = \left(\frac{\rho_\odot}{m_{\chi}}\right) \int du \hspace{0.5mm} \frac{f(u)}{u} 
\int_0^{R_\odot} dr \hspace{0.5mm} 4 \pi r^2 \hspace{0.6mm} w^2 
\hspace{0.5mm} \sum_{T} 
\Theta\left(w^2 - v^2_{T*}\right) 
\hspace{1mm} \Omega_T ,
\label{eq:cap_rate}
\end{equation}
with 
\begin{equation}
\Omega_T = \eta_{T}(r) \hspace{0.5mm} \hspace{0.5mm} \Theta(E^\chi_{\rm max}-E^\chi_{\rm cap}) 
\int_{{\rm max}[E^\chi_{\rm min},E^\chi_{\rm cap}]}^{E^\chi_{\rm max}} dE \hspace{1mm} \frac{d\sigma_T}{dE} .
\label{eq:interaction_rate}
\end{equation}
Here $f(u)$ is the normalised local speed distribution of WIMPs (with respect to the solar reference frame) 
with $u$ the WIMP speed far away from the Sun, 
and $\rho_\odot$ is the local DM density for which we use the standard value 
$\rho_\odot = 0.3$ $\rm GeV/cm^{3}$~\cite{DD_LEWIN1996}. 
Inside the gravitational field of the Sun WIMPs accelerate to a larger speed $w$: 
\begin{equation}
w(u,r) = \sqrt{u^2 + v_{\rm esc}(r)^2}
\end{equation}
where $v_{\rm esc}(r)$ is the local escape velocity at a radius $r$, varying in the range 620--1400 km/s from the surface to the center. 
For the number density profile $\eta_{T}(r)$ of the different nuclear targets 
in the Sun we use the Standard Solar Model AGSS09ph~\cite{solar_model_Serenelli2009}.
The minimum (maximum) energy $E^\chi$ deposited by the WIMP in the inelastic scattering process, and 
the minimum energy that the WIMP needs to lose in order to be captured are, respectively: 
\begin{eqnarray}
E^\chi_{\rm min, \rm max}(w) &=& \frac{1}{2} m_{\chi} w^2 \left[1 - \frac{\mu^2_{\chi T}}{m^2_T} 
\left(1 \pm \frac{m_T}{m_{\chi}} 
\sqrt{1 - \frac{v^2_{T*}}{w^2}}\right)^2 \right] - \delta ,
\label{eq:E_max_min}
\\
E^\chi_{\rm cap}(u) &=& \frac{1}{2} m_{\chi} u^2 - \delta .
\label{eq:E_cap}
\end{eqnarray}

Capture of a WIMP in the Sun through the scattering process off a target $T$ is kinematically possible up to  
a maximum value of the asymptotic WIMP speed ($u^{\rm C-max}_T$) which is determined from the condition: 
\begin{eqnarray}
E^\chi_{\rm max} &\geq& E^\chi_{\rm cap}, 
\hspace{3mm} {\rm or},  \hspace{3mm}
w^2 \left[1 - \frac{\mu^2_{\chi T}}{m^2_T} 
\left(1 - \frac{m_T}{m_{\chi}} 
\sqrt{1 - \frac{v^2_{T*}}{w^2}}\right)^2 \right] \geq u^2 .
\label{eq:E_max_min_condition}
\end{eqnarray} 

The equation above is equivalent to requiring that the outgoing speed of the heavy state $\chi^{\prime}$ is below the escape velocity $v_{\rm esc}(r)$ at the position where the WIMP--nucleus scattering process takes place inside the celestial body. We assume that the heavier state $\chi^\prime$ produced in the inelastic scattering $\chi + T \rightarrow \chi^\prime + T$ is short--lived and quickly decays back to the DM particle $\chi$ along with other much lighter particle(s) which carry away most of the energy produced in the decay, so that the outgoing $\chi$ does not receive a significant amount of kinetic energy and remains locked in a bound orbit. As a consequence it is possible to assume that the DM particle is captured after a single scattering event complying with Eq.~(\ref{eq:E_max_min_condition}) so that the optically thin approximation of Eqs.~(\ref{eq:cap_rate}, \ref{eq:interaction_rate}) can be used to estimate the capture rate. 

After being locked into bound orbits, during the lifetime of the solar system the DM particles continue to scatter off the nuclear targets in the Sun and are eventually driven to its core, where their annihilation process can be probed by Neutrino Telescopes (NTs). In particular in the case of elastic scattering the WIMPs continue to scatter until they thermalize with the nuclei of the solar plasma, developing a population with density profile:

\begin{eqnarray}
n_\chi(r) &=&   n_0 e^{-m_\chi\phi(r)/T_c} \simeq n_0 e^{-r^2/r_\chi^2}\nonumber\\
r_\chi^2&=&\frac{3 k_B T_c}{2\pi G\rho_c m_\chi},
\label{eq:sun_wimp_density}
\end{eqnarray}

\noindent with $n_0$ a normalization constant, $k_B$ the Boltzmann constant, $\rho_c\simeq$ 150 g/cm$^3$ the central density of the Sun and $T_c\simeq 1.55\times 10^7$ $^{\circ}$K the central temperature~\cite{griest_seckel_1986}, while:

\begin{equation}
    \phi(r)=\int_0^r \frac{G M(r)}{r^2}\,dr,
\end{equation}

 \noindent is the gravitational potential inside the Sun. It has been recently pointed out that this assumption may not be robust in the case of inelastic scattering, because, depending on the value of the mass splitting $\delta$, the WIMPs can eventually stop interacting with the nuclei in the Sun when their dispersion velocity is too low to trigger the inelastic process~\cite{Blennow_2018}. 
In Appendix~\ref{app:thermalization} we show that even when the quantity $C_A$ defined in (\ref{eq:ca}) is taken to its smallest possible value (irrespective of thermalization) the effect on the constraints is relatively mild, and, in particular, does not affect the results of the discussion in Section~\ref{sec:analysis}. As a consequence, for definiteness in the following we will present our results assuming thermalization and the standard expression of Eq.~(\ref{eq:sun_wimp_density}) for the WIMP density within the Sun, for which:

\begin{eqnarray}
    C_A&=&\langle \sigma v \rangle \frac{V_2}{V_1^2}\nonumber\\
    V_n&=&4\pi\int_o^{r_\odot} r^2 e^{-n r^2/r_\chi^2}\,dr.
    \label{eq:ca_thermalization}
\end{eqnarray}

\subsection{Direct detection}
\label{sec:dd}

The total number of expected events induced by the WIMP--nucleus inelastic scattering process
in a DD experiment can be written as:
\begin{eqnarray}
R_{\rm DD} &=& M{\tau_{\rm exp}} \hspace{0.5mm} \left(\frac{\rho_\odot}{m_{\chi}}\right) 
\int du \hspace{0.5mm} f(u) \hspace{0.5mm} u \hspace{0.5mm} 
\sum_{T} N_{T} \hspace{0.6mm} \Theta\left(u^2 - v^2_{T*}\right)
\int_{E_{\rm min}(u)}^{E_{\rm max}(u)} dE \hspace{0.6mm} \hspace{0.5mm} 
\zeta_T \hspace{1mm} \frac{d\sigma_T}{dE} ,
\label{eq:DD_event}
\end{eqnarray}
where $M$ is the fiducial mass of the detector, 
$\tau_{\rm exp}$ is the live--time of data taking and 
$N_T$ is the number of targets per unit mass in the detector. 
The minimum (maximum) recoil energy of the nucleus $E_{\rm min, max} (u) = E_{\rm min, max} (w=u)$ 
are given by:

\begin{equation}
     E_{\rm min,max}(u)=\frac{\mu^2_{\chi T}u^2}{2 m_T}\left ( 1\mp\sqrt{1-\frac{2\delta}{\mu_{\chi T}u^2}}  \right )^2.
  \label{eq:er_max_min}
\end{equation}

In Eq.~(\ref{eq:DD_event}) $u$ indicates the incoming WIMP speed in the reference frame of the Earth, approximated to the solar reference frame (i.e. neglecting, or averaging away the Earth rotation around the Sun\footnote{In the original paper~\cite{Halo-independent_Ferrer2015}, in order to obtain the most conservative bound including the effect of the Earth rotation around the Sun the authors minimized the rate with respect to the angle between the stream velocity $u$ and a fixed Earth velocity $v_E$. To make the discussion in the following Sections more transparent we choose to neglect the effect of the Earth rotation around the Sun altogether (i.e. we set $v_E$ = 0) because we find that it is mild on the DD-NT combined bounds (unless for large WIMP masses, where the maximal value of $\delta$ for which a halo--independent bound is possible can be reduced by $\lesssim$ 30\%).}). Moreover, the function $\zeta_T$ in Eq.~(\ref{eq:DD_event}) indicates the response of the detector which can be expressed in terms of the experimental energy bin $[E_1, E_2]$ and the efficiency $\epsilon(E)$ as:
\begin{equation}
\zeta_T = \epsilon(E) \Theta(E-E_1) \Theta(E_2-E) .
\label{eq:zt}
\end{equation}

A direct detection experiment (with target $T$) is sensitive only to the 
speed range $u \geq u^{\rm DD-min}_T$ with:

\begin{numcases}{u^{\rm DD-min}_T=}
& $v_{T*}$ = $\sqrt{\frac{2\delta}{\mu_{\chi T}}}$,\,\,\,\,\,\,\,\,\,\, $E_T^{\rm th}\leq E_{T*}$\nonumber\\
& $\frac{1}{\sqrt{2 m_T E_T^{\rm th}}}\left ( \frac{m_T E_T^{\rm th}}{\mu_{\chi T}}+\delta\right)$,\,\, 
$E_T^{\rm th}> E_{T*}$ \label{eq:uth}
\end{numcases}

\noindent with $E^{\rm th}_T$ the recoil energy threshold and $E_{T*}$ = $\mu_{\chi T}\delta/m_T$.
In Eq~(\ref{eq:DD_event}) $E^{\rm th}_T$ enters through the function 
$\zeta_T$ which is zero below $E^{\rm th}_T$.

\section{Experimental sensitivity to the IDM kinematics}
\label{sec:combined_sensitivity_idm}

\begin{figure*}[ht!]
\centering
\includegraphics[width=7.49cm,height=6cm]{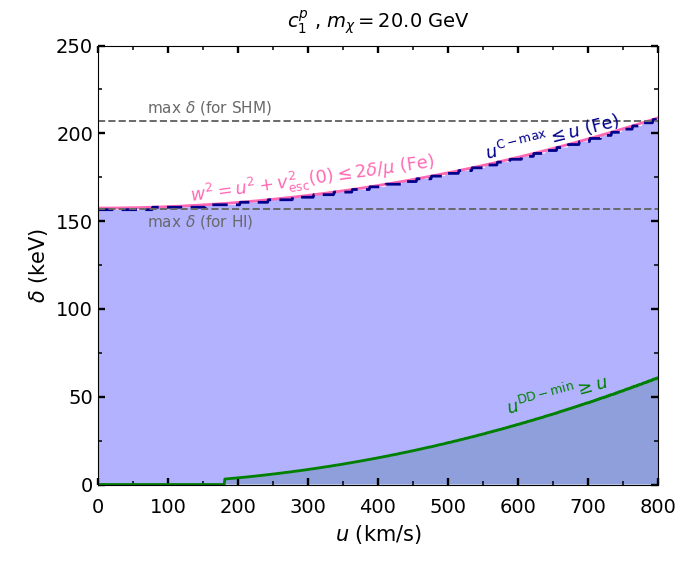}
\includegraphics[width=7.49cm,height=6cm]{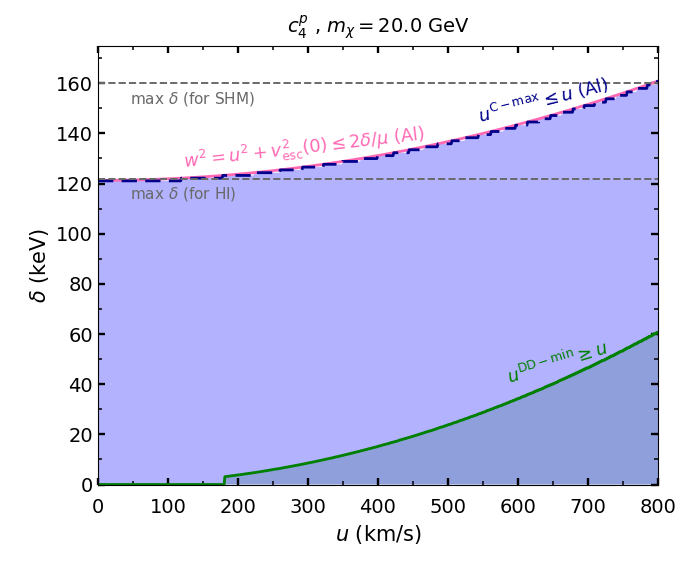}
\caption{Sensitivities of DD and capture in the Sun to the range of the WIMP speed $u$ shown in the $u$ -- $\delta$ plane for $m_\chi$ = 20 GeV. Left--hand plot: SI interaction; right--hand plot: SD interaction. In each figure the blue (green) shaded region corresponds to capture in the Sun (DD). The overall sensitivity achievable by combining DD and Capture is the union of the two regions (in this specific case the green DD region is contained in the blue one, i.e. capture in the Sun provides alone the halo--independent bound when the latter is possible). }
\label{fig:delta_u_mchi_20_GeV}
\end{figure*}

\begin{figure*}[ht!]
\centering
\includegraphics[width=7.49cm,height=6cm]{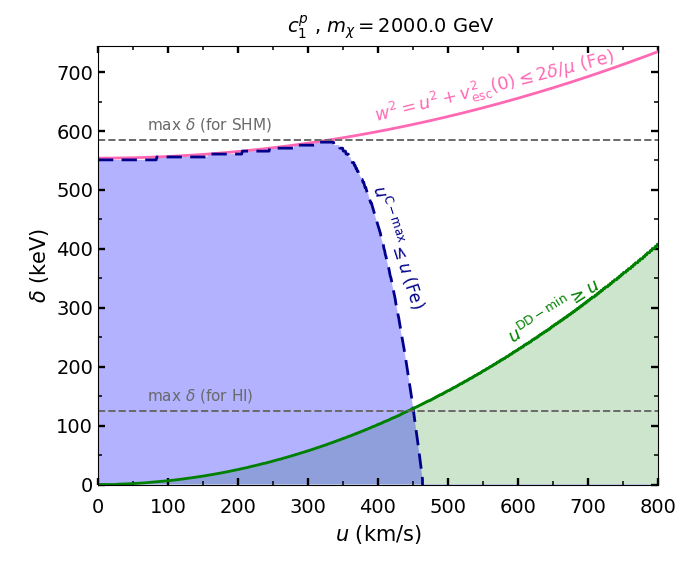}
\includegraphics[width=7.49cm,height=6cm]{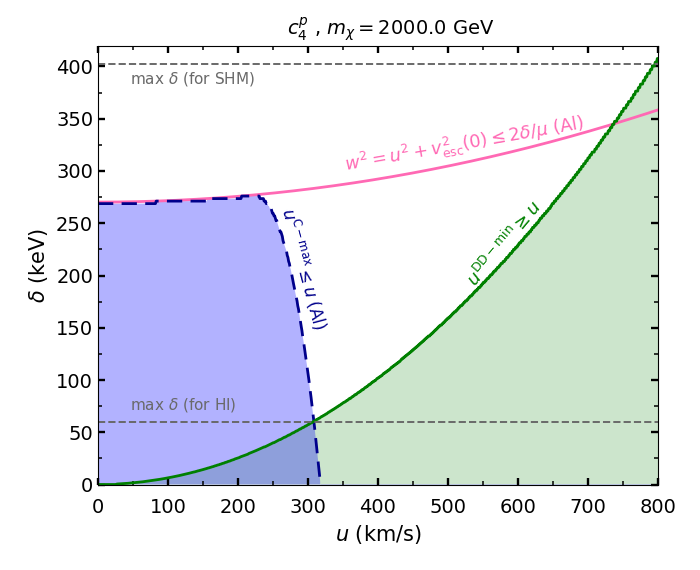}
\caption{The same as in Fig.~\ref{fig:delta_u_mchi_20_GeV} for $m_\chi$ = 2 TeV.}
\label{fig:delta_u_mchi_2_TeV}
\end{figure*}

In the following Sections we assume that the local distribution of the WIMP speed with respect to the solar reference frame is described by some function $f(u)$ normalised as:
\begin{equation}
\int^{u_{\rm max}}_0 du \hspace{0.5mm} f (u) = 1 ,
\label{eq:f_u_norm}
\end{equation}
where $u$ is the WIMP speed far away from the Sun and $u_{\rm max}$ is its 
maximum possible value. Assuming that all the DM particles in the halo are gravitationally bound to the Galaxy, 
$u_{\rm max} = u_{\rm esc} + v_\odot$, with $u_{\rm esc}$ the escape speed at the solar position. In our analysis we will take the reference value $u_{\rm esc} = 560$ $\rm km/s$~\cite{vesc_Smith2006, vesc_Piffl2013} implying $u_{\rm max}$ = $u_{\rm max}^{ref}$ = 780 $\rm km/s$. In appendix~\ref{app:umax_dep} we discuss in detail the consequences of assuming $u_{\rm max}> u_{\rm max}^{ref}$.

The discussion of Section~\ref{sec:idm_scattering} singles out specific kinematic conditions in order for the DD process and/or for capture in the Sun measured by NT's to take place. Namely, we summarize them here for convenience:
\begin{eqnarray}    
&& u^2+v_{\rm esc}^2(r=0)>v_{T*}^2,\,\,\mbox{(NT)} \label{eq:kin1}\\
&& u>u^{\rm DD-min},\,\,\mbox{(DD)}\label{eq:kin2}\\
&& u<u^{\rm C-max},\,\,\mbox{(NT)} \label{eq:kin3}\\
&& u^{\rm DD-min}<u^{\rm C-max},\,\,\mbox{(DD+NT)} \label{eq:kin4}
\end{eqnarray}

\noindent where $v_{T*}^2$ and $u^{\rm DD-min}$ are given in Eqs.~(\ref{eq:vstar}, \ref{eq:uth}), while $u^{\rm C-max}$  is obtained from Eq.~(\ref{eq:E_max_min_condition}).
For a given choice of the WIMP mass $m_\chi$ and $\delta$ such conditions determine the range of the WIMP incoming velocity $u$ to which DD and NT are sensitive. In particular~(\ref{eq:kin1}) is the condition for the inelastic scattering process to be kinematically possible, (\ref{eq:kin2}) for the recoil energy deposited in the scattering process to be above the DD experimental threshold and (\ref{eq:kin3}) for the outgoing speed of the WIMP to be below the escape velocity within the Sun in capture. Conditions~(\ref{eq:kin2}) and (\ref{eq:kin3}) are valid in specific intervals of $u$. In particular, at low--enough values of the WIMP mass $m_\chi$ the condition~(\ref{eq:kin3}) for capture can encompass the full range of $u$, $0\le u\le u_{\rm max}$, allowing to obtain a halo--independent bound from capture in the Sun alone, while at higher WIMP masses the two conditions~(\ref{eq:kin2}) and (\ref{eq:kin3}) must be verified at the same time, i.e. condition~(\ref{eq:kin4}) must hold.

For a given value of $m_\chi$ the conditions (\ref{eq:kin1}--\ref{eq:kin4}) single out two regions in the $u$ -- $\delta$ plane, one for DD and the other for NT, the {\it union} of the two regions providing the overall experimental sensitivity to $u$. Such regions are provided in Figs.~\ref{fig:delta_u_mchi_20_GeV} and \ref{fig:delta_u_mchi_2_TeV} for $m_\chi$ = 20 GeV and 2 TeV, respectively. For the experimental setups considered in Section~\ref{sec:analysis}, in the case of a SI interaction the boundaries are determined by $T=Xe$ for DD and $T=Fe$ for NT, while for a SD interaction they are driven by $T=Xe$ or $T=I$ for DD and $T=Al$ for NT.

In particular, in such figures the range of $u$ sampled by DD and NT for a given value of $\delta$ corresponds to the part of the horizontal line at constant $\delta$ that overlaps with the union of the green and blue shaded regions. From such figures one can immediately notice that two different scenarios arise at small and large $m_\chi$. At low--enough values of $m_\chi$ (as in the case $m_\chi$ = 20 GeV shown in Fig.~\ref{fig:delta_u_mchi_20_GeV}) the green--shaded region of sensitivity to DD is completely contained inside the blue--shaded area for NT. This indicates that in this case the NT bound is sufficient alone to provide a halo--independent constraint for those values of $\delta$ for which the blue shaded area extends in the full range of $u$, i.e. 0$\le u\le u_{\rm max}$. This turns out to be possible as long as $v^2_{T*}<u^2+v_{\rm esc}^2(r=0)$ with $T$ = $Fe$ for a SI interaction and $T$ = $Al$ for a SD one (for $m_\chi$ = 20 GeV this corresponds in Fig.~\ref{fig:delta_u_mchi_20_GeV} to $\delta\lesssim$ 150 keV for a SI interaction and $\delta\lesssim$ 120 keV for a SD one), i.e. as long as the inelastic process is kinematically allowed in the first place.  This is confirmed by the fact that in Fig.~\ref{fig:delta_u_mchi_20_GeV} the boundary of the region allowed by capture has the same dependence on 
$\delta$ than that for DD, since in both cases it is given by the condition that the inelastic process is kinematically accessible, albeit for a different maximal initial speed of the WIMP.
On the other hand, the example of Fig.~\ref{fig:delta_u_mchi_2_TeV} for $m_\chi$ = 2 TeV shows that at higher values of $m_\chi$ the extension of the blue shaded area corresponding  to NT no longer reaches $u_{\rm max}$, and the green region corresponding to DD becomes crucial to fill the gap, as long as 
$\delta$ is small enough. In this case a halo independent bound is possible if capture in the Sun and DD  are complementary in sampling the full range, i.e. if $\delta$ lies below the lower dashed horizontal line that passes through the intersection of the blue dashed line and the green solid one (i.e. if $\delta\lesssim$ 120 keV for a SI interaction and $\delta\lesssim$ 60 keV for a SD one). In this case inspection of Fig.~\ref{fig:delta_u_mchi_2_TeV} shows that, although the kinematic range of $u$ accessible to DD and that for capture are both reduced when $\delta$ gets larger, they do in different ways. Namely, increasing $\delta$ pushes $u^{\rm DD-min}$ for DD to larger values and $u^{\rm C-max}$ to lower values, but the dependence of  $u^{\rm C-max}$ on $\delta$ is much milder than that of $u^{\rm DD-min}$. In other words, in this regime ($m_\chi>>m_T$) the inelasticity of the scattering process reduces the experimental sensitivity to capture in the Sun to a lesser extent compared to DD. In order to understand this one can notice that the sensitivity of DD is determined by the maximal value of $E_R$  which, indicating with $N$ the corresponding nuclear target, is given by Eq.~(\ref{eq:er_max_min}):

\begin{equation}
E_R^{\rm max}=\frac{\mu_{\chi N}^2 u^2}{2 m_N}\left (1+\sqrt{1-\frac{2\delta}{\mu_{\chi N} u^2}} \right )^2 \,.
\label{eq:e_max_dd}
\end{equation} 

On the other hand the sensitivity of capture on the target $T$ in the Sun corresponds to the minimal value $E^\chi_{f,\rm min}$ of $E^{\chi}_{f}$ (corresponding to the outgoing speed of $\chi^{\prime}$), which can be obtained from the expression of $E^{\chi}_{\rm max}$ of Eq.~(\ref{eq:E_max_min}):

\begin{equation}
E^\chi_{f,\rm min}=\frac{m_\chi}{m_T}\frac{\mu_{\chi T}^2 w^2}{2 m_T}\left (1-\frac{m_T}{m_\chi}\sqrt{1-\frac{2\delta}{\mu_{\chi T} w^2}} \right )^2.
\label{eq:e_min_NT}
\end{equation} 

In particular when $m_T/m_\chi\ll 1$ the expressions of Eq.~(\ref{eq:e_min_NT}) has a much milder dependence on $\delta$ compared to that of Eq.~(\ref{eq:e_max_dd}).

Irrespective on the WIMP mass if $\delta$ is too large to provide a complete sampling of the 
full range 0$<u<u_{\rm max}$, a partial sampling is sufficient to obtain a bound if the velocity distribution $f(u)$ is fixed (for instance, if a Maxwellian is adopted). For larger values of $\delta$ eventually capture in the Sun becomes kinematically impossible for any value of $u$ and no bound is possible even fixing $f(u)$ (in Figs.~\ref{fig:delta_u_mchi_20_GeV} and \ref{fig:delta_u_mchi_2_TeV} this corresponds to the value of $\delta$ for which the sampled range of $u$
shrinks to a point, i.e. to the upper dashed horizontal line).  For even larger values of $\delta$ the velocity $v_{T*}$  exceeds $u_{\rm max}$ and the WIMP inelastic scattering process off the target $T$ becomes kinematically non--accessible in the first place. 

By comparing Fig.~\ref{fig:delta_u_mchi_20_GeV} for $m_\chi$ = 20 GeV to Fig.~\ref{fig:delta_u_mchi_2_TeV} for $m_\chi$ = 2 TeV one can notice that increasing $m_\chi$ the different kinematic domains are simply shifted to higher values of $u$. In particular the same pattern of domain shapes observed in  Fig.~\ref{fig:delta_u_mchi_20_GeV} for 0$\le u\le u_{\rm max}$ with $u_{\rm max}\simeq$ 800 km/sec can be recognized in Fig.~\ref{fig:delta_u_mchi_2_TeV} for 0$\le u\lesssim$ 350 km/s in the case of a SI interaction, and for 0$\le u\lesssim$ 250 km/s in the case of a SD one.

\section{Halo--independent bounds with the single--stream method}
\label{sec:single_stream}

In the previous Section we have discussed the kinematic conditions that allow to obtain a halo--independent bound by combining DD and Capture in the Sun.
In this Section we outline the procedure originally introduced in~\cite{Halo-independent_Ferrer2015} to quantitatively determine such bound when it is kinematically possible. 

The capture rate of WIMPs in the Sun (Eq.~(\ref{eq:cap_rate})) or the expected number of WIMP--induced nuclear recoil events in a 
direct detection experiment (Eq.~(\ref{eq:DD_event})) can be written as:
\begin{equation}
R = \int^{u_{\rm max}}_0 du \hspace{0.5mm} f(u) \hspace{0.5mm} H(u) .
\label{eq:rate_H_u}
\end{equation}
For capture, 
\begin{eqnarray}
H(u) = H_C(u) &=& \left(\frac{\rho_\odot}{m_{\chi}}\right) \frac{1}{u} 
\int_0^{R_\odot} dr \hspace{0.5mm} 4 \pi r^2 \hspace{0.6mm} w^2 
\hspace{0.5mm} \sum_{T} 
\Theta\left(w^2 - v^2_{T*}\right) 
\nonumber\\ && \times 
\hspace{0.5mm} \eta_{T}(r) \hspace{0.5mm} \hspace{0.5mm} \Theta(E^\chi_{\rm max}-E^\chi_{\rm cap}) 
\int_{{\rm max}[E^\chi_{\rm min},E^\chi_{\rm cap}]}^{E^\chi_{\rm max}} dE \hspace{1mm} \frac{d\sigma_T}{dE} ,
\label{eq:NT_H_u} 
\end{eqnarray}
and for direct detection, 
\begin{eqnarray}
H(u) = H_{\rm DD}(u) &=& M{\tau_{\rm exp}} \hspace{0.5mm} \left(\frac{\rho_\odot}{m_{\chi}}\right) 
u \hspace{0.5mm} 
\sum_{T} N_{T} \hspace{0.6mm} \Theta\left(u^2 - v^2_{T*}\right)
\int_{E_{\rm min}(u)}^{E_{\rm max}(u)} dE \hspace{0.6mm} \hspace{0.5mm} 
\zeta_T \hspace{1mm} \frac{d\sigma_T}{dE} .
\label{eq:DD_H_u}
\end{eqnarray}

Considering a WIMP--nucleon coupling, say $c$, a given experimental bound $R_{\rm max}$ (either from a DD experiment or a NT) implies:
\begin{equation}
R = R(c^2) = \int^{u_{\rm max}}_0 du f(u) H(c^2, u) \leq R_{\rm max} ,
\label{eq:rate1}
\end{equation}
with $H$ being either $H_C$ or $H_{\rm DD}$. Since $H(c^2, u) = c^2 H(c = 1, u)$, one can re--write (\ref{eq:rate1}) as:
\begin{equation}
R(c^2) = \int^{u_{\rm max}}_0 du f(u) \frac{c^2}{{c^2}_{\rm max}(u)} H({c^2}_{\rm max}(u), u) = 
\int^{u_{\rm max}}_0 du f(u) \frac{c^2}{{c^2}_{\rm max}(u)} R_{\rm max} \leq R_{\rm max} ,
\label{eq:rate2}
\end{equation}
where ${c}_{\rm max}(u)$ is defined as:
\begin{equation}
H({c^2}_{\rm max}(u), u) = {c^2}_{\rm max}(u) H(c=1, u) = R_{\rm max} ,
\label{eq:c_max_u}
\end{equation}
i.e., ${c}_{\rm max}(u)$ is the upper limit on the coupling $c$ if all the incoming WIMPs 
are in a single stream with speed $u$. From Eq.~(\ref{eq:rate2}) one obtains the following upper limit on the 
coupling $c$ for a general WIMP speed distribution $f(u)$:
\begin{equation}
c^2 \leq \left[\int^{u_{\rm max}}_0 du \frac{f(u)}{{c^2}_{\rm max}(u)}\right]^{-1} .
\label{eq:c2_upper_bound}
\end{equation}


A conservative bound independent of $f(u)$ on the coupling $c$ can then be obtained if in the expression above a finite maximum $\tilde{c}$ over the full range of $u$ of the quantity ${c^2}_{\rm max}(u)$ can be factored out from the integral. This is possible if the two following conditions are satisfied:

\begin{enumerate}
\item $w(u$=0$, r$=0$) \geq v_{*}$ (at least for a few dominant targets in the Sun), 
so that the WIMP capture rate is not zero (or negligible) for small values of 
the WIMP speed $u$ to which direct detection experiments are not sensitive.


\item $u^{\rm C-max} \geq$ min$\left[u^{\rm DD-min}, u_{\rm max}\right]$ 
(at least for a few dominant targets in the Sun), 
so that either neutrino telescopes alone or neutrino telescopes and direct detection experiments together 
can effectively probe the full WIMP speed range $[0, u_{\rm max}]$.
\end{enumerate}

In particular, {\it if any of the two above--mentioned conditions is not satisfied a halo--independent bound is not possible}. On the other hand, if both conditions are true one of the two following situations may occur (see Fig.~\ref{fig:ci_u_diff_delta} for specific examples):  

\begin{itemize}
\item \underline{Case A:}
The NT and the DD experiments are sensitive to two specific parts of the full WIMP speed range, 
giving:
\begin{eqnarray}
{({c^{\rm NT}})^2}_{\rm max}(u) &\leq& \tilde{c}^2 
\hspace{18mm} {\rm for} \hspace{2mm} 0 \leq u \leq \tilde{u}
\nonumber\\
{({c^{\rm DD}})^2}_{\rm max}(u) &\leq& \tilde{c}^2 
\hspace{18mm} {\rm for} \hspace{2mm} \tilde{u} \leq u \leq u_{\rm max}
\label{eq:u_tilde_c_tilde}
\end{eqnarray}
where ${{c^{\rm NT}}}_{\rm max}(u)$ and ${{c^{\rm DD}}}_{\rm max}(u)$ 
correspond to ${c}_{\rm max}(u)$ for the NT and the DD experiments, respectively, 
and $\tilde{u}$ denotes the speed where 
${{c^{\rm NT}}}_{\rm max}$ and ${{c^{\rm DD}}}_{\rm max}$ intersect at a finite value $\tilde{c}$, 
i.e., ${{c^{\rm NT}}}_{\rm max}(\tilde{u}) = {{c^{\rm DD}}}_{\rm max}(\tilde{u}) = \tilde{c}$. 
In this case, using Eq.~(\ref{eq:c2_upper_bound}) it can be shown that 
(see \cite{Halo-independent_Ferrer2015} and \cite{halo_independent_sogang_2023} for details) 
the halo--independent bound on the coupling is:
\begin{equation}
c^2 \leq 2 \hspace{0.5mm} \tilde{c}^2 .
\label{eq:limit_A}
\end{equation}

\item \underline{Case B:}
The NT constraint is stronger than that from DD in the full speed range $[0, u_{\rm max}]$, i.e., 
\begin{equation}
{{c^{\rm NT}}}_{\rm max}(u) < {{c^{\rm DD}}}_{\rm max}(u) 
\hspace{3mm} {\rm for} \hspace{2mm} u \in [0, u_{\rm max}] . 
\end{equation}
In this case, the halo--independent bound on the coupling is simply: 
\begin{equation}
c^2 \leq {\rm max} \left[ {({c^{\rm NT}})^2_{\rm max}} \left(u\right) \right] . 
\label{eq:limit_B}
\end{equation}
\end{itemize}

For a given WIMP--nucleon interaction, at each $m_{\chi}$ and $\delta$ we calculate the 
halo--independent upper limit on the interaction coupling following 
either Eq.~(\ref{eq:limit_A}) or (\ref{eq:limit_B}), when appropriate.
When more than one DD and one NT are involved (as in our analysis) 
the procedure described above is repeated by combining each 
DD bound with each NT bound, and the most constraining halo--independent limit on the coupling is taken.

\section{Analysis}
\label{sec:analysis}

\begin{figure*}[ht!]
\centering
\includegraphics[width=7.49cm,height=6cm]{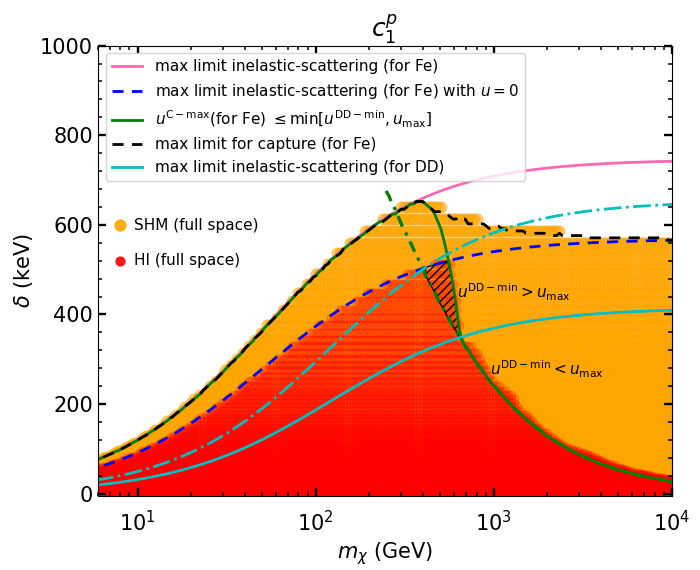}
\includegraphics[width=7.49cm,height=6cm]{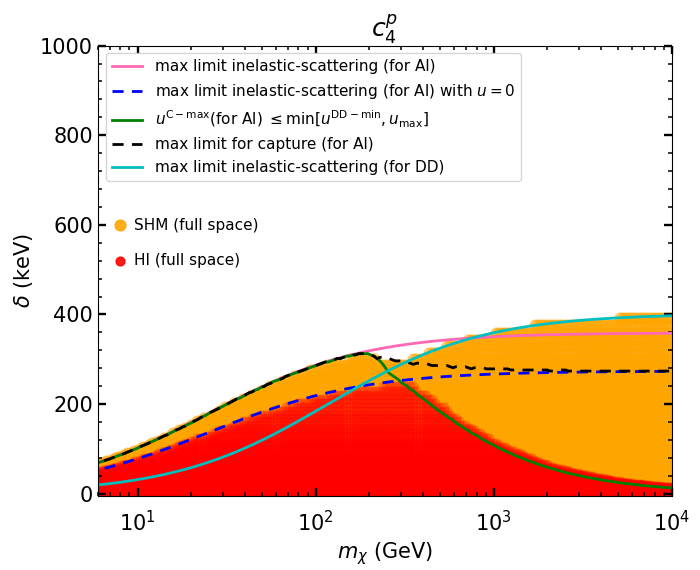}
\caption{$m_\chi$ -- $\delta$ parameter space of Inelastic Dark Matter for which a single--stream halo--independent upper bound on the coupling is possible (red--shaded region) or an upper bound for the Standard Halo Model is possible (extended by the orange--shaded region). {\bf Left--hand plot:} spin--independent interaction. The dashed area represents the reduction of the experimental sensitivity to $\delta$ when $u_{\rm max}^{ref}<u_{\rm max}<u_{\rm max}^{\rm DD-C}\simeq$ 980 km/s (see Section~\ref{sec:analysis} and Appendix~\ref{app:umax_dep}). The dashed light--green line represents $\delta_{\rm max, DD}$ for $u_{\rm max}$ = $u_{\rm max}^{\rm DD-C}$. Taking $u_{\rm max}>u_{\rm max}^{\rm DD-C}$ does not change the red--shaded region further.  {\bf Right--hand plot:} Spin--dependent interaction. In 
both plots the cut $c \le g^2/M^2$ with $g = \sqrt{4\pi}$ and $M = 1$ GeV is applied. Changing this cut by one order of magnitude in both directions does not modify the shaded regions significantly. 
}
\label{fig:delta_mx_scan}
\end{figure*}

For definiteness in this Section we will assume thermalization of the WIMPs with the solar plasma and fix $u_{\rm max}$ = $u_{\rm max}^{ref}$ = 780 km/s. See Appendix~\ref{app:thermalization} for a discussion of the consequences of relaxing the thermalization assumption, and Appendix~\ref{app:umax_dep}    for those of assuming $u_{\rm max}> u_{\rm max}^{ref}$. 

The main results of our analysis are shown in Figs.~\ref{fig:delta_mx_scan}
and Fig.~\ref{fig:limit_mx_diff_delta}. In particular in Fig.~\ref{fig:delta_mx_scan} the red--shaded regions indicate the values of $m_\chi$ and $\delta$ for which a single--stream halo--independent bound obtained following the procedure outlined in Section~\ref{sec:single_stream} is possible. In the same figure the red--shaded areas are extended by the orange--shaded ones when the velocity distribution $f(u)$ is fixed to the Standard Halo Model, i.e. to a Maxwellian with velocity dispersion $u_{\rm rms}=270$ km/s~\cite{SHM_maxwell_Green2011}, cut at the escape velocity $u_{\rm esc} = 560$ $\rm km/s$~\cite{vesc_Smith2006, vesc_Piffl2013} 
and boosted by the rotational speed of the solar system $v_\odot = 220$ $\rm km/s$~\cite{SHM_maxwell_Green2011}.

In such figure the left--hand plot corresponds to the SI case and the right--hand plot to the SD one. On the other hand, in Fig.~\ref{fig:limit_mx_diff_delta} the exclusion plots of the SI coupling $c_1$ and for the SD coupling $c_4$ are provided as a function of the WIMP mass $m_\chi$ for some representative values of the mass splitting parameter $\delta$.

As discussed in Section~\ref{sec:combined_sensitivity_idm} the conditions that determine the shaded regions of Fig.~\ref{fig:delta_mx_scan} are mainly due to kinematics. Technically, this would imply a diverging bound on the edges of such domains. However, a stronger condition than divergence of the bound can be obtained by imposing a cut corresponding to the validity of the effective theory and on perturbativity. In particular, for a contact interaction one can express the dimensional effective coupling  as $c = g^2/M^2$, and impose i) the perturbativity bound $g\le \sqrt{4\pi}$; ii) that the scale $M$ is larger than the nucleon mass, i.e. $M\gtrsim$ 1 GeV. In Fig.~\ref{fig:delta_mx_scan} we follow this procedure applying the cut $c\lesssim 4\pi$ GeV$^{-2}$ =  12.6 GeV$^{-2}$ rather than $c<\infty$. Changing this cut by one order of magnitude in both directions does not modify the shaded regions 
significantly\footnote{We checked that the typical values of the transferred momentum $q$ both in the case of direct detection experiments and for Capture in the Sun is less than a GeV. Thus, it is possible to assume an effective contact interaction even for a mediator mass scale $\mathcal{O}(1)$ GeV.}.

The different domains of Fig.~\ref{fig:delta_mx_scan} can be easily understood in terms of the kinematic  discussion provided in Section~\ref{sec:combined_sensitivity_idm} and looking at Figs.~\ref{fig:delta_u_mchi_20_GeV} and \ref{fig:delta_u_mchi_2_TeV}. In particular, it is instructive to focus on the maximal values $\delta_{\rm max}$ of the halo--independent bound  on $\delta$ (i.e. the maximal value of $\delta$ that can be probed in a halo--independent way) comparing the two cases of Fig~\ref{fig:delta_mx_scan} for a SI and a SD interaction. 

\begin{figure}
    \centering
    \includegraphics[width=4.9cm]{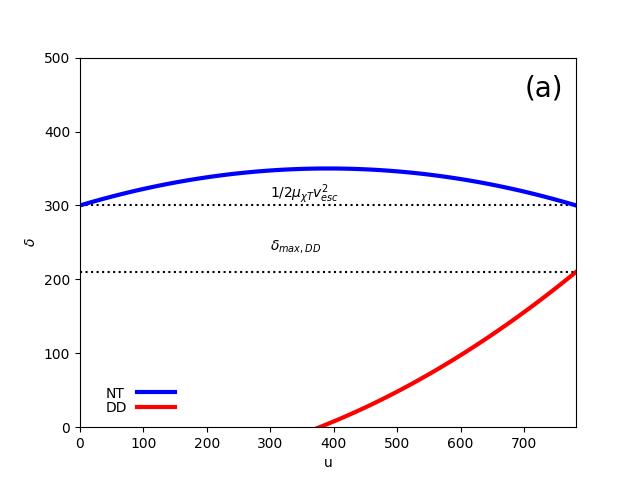}
    \includegraphics[width=4.9cm]{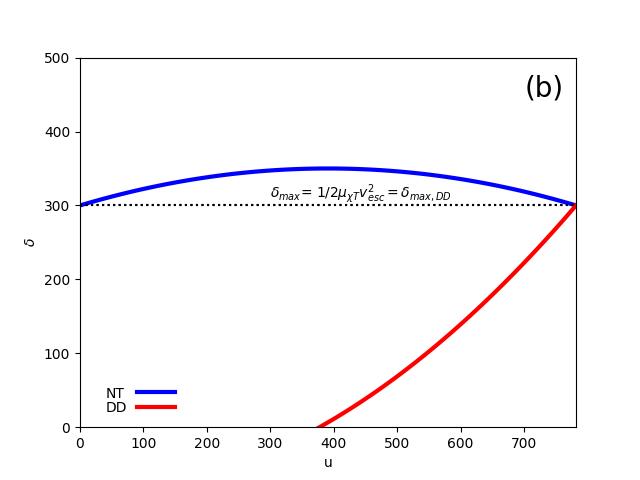 }
    \includegraphics[width=4.9cm]{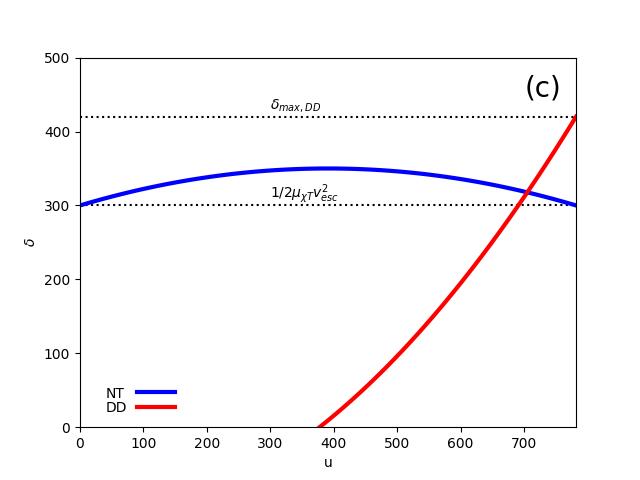 }
    \caption{
    Schematic exemplification on why $\delta_{\rm max,DD}$, the maximal value of $\delta$ probed by DD, needs to be larger than the maximal value of $\delta$ probed by capture for $u$ =0 ($\delta=1/2\mu_{\chi T} v_{\rm esc}^2$) in order for DD to play a role in the determination of $\delta_{\rm max}$, the maximal value of $\delta$ that can be constrained in a halo--independent way.
    The plots correspond to three increasing values of the mass $m_N$ of the DD target $N$ and in each of them the $u$ -- $\delta$ parameter space below the blue solid line is probed by capture in the sun, while that below the red solid lines is probed by DD. {\bf (a):} $\delta_{\rm max}$ is only determined by capture on the target $T$ within the Sun because when the two conditions $v_{\rm esc}$ = $v_{T*}$ and $u^{\rm C-max}$ = $u_{\rm max}$ are verified one has $u^{\rm DD-min}>u_{\rm max}$ and the DD signal vanishes;  {\bf (c):} when $v_{\rm esc}$ = $v_{T*}$ one also has $u^{\rm DD-min}<u_{\rm max}$; {\bf (b):} the boundary between the two situation corresponds to $\delta_{\rm max, DD}$ = 1/2 $\mu_{\chi T} v_{\rm esc}^2$ and allows to determine the minimal value of $m_N$ that the DD target needs to have in order to compete with capture in the Sun in the determination of 
    $\delta_{\rm max}$ (see Eqs.~(\ref{eq:vnstar1}) and (\ref{eq:mn_min})).
    }
    \label{fig:u_delta_max}
\end{figure}

\begin{figure*}[ht!]
\centering
\includegraphics[width=7.49cm,height=6cm]{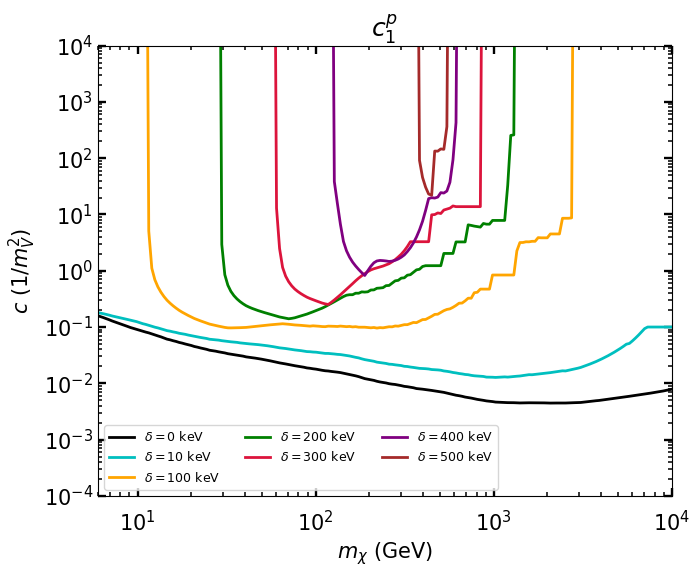} \\
\includegraphics[width=7.49cm,height=6cm]{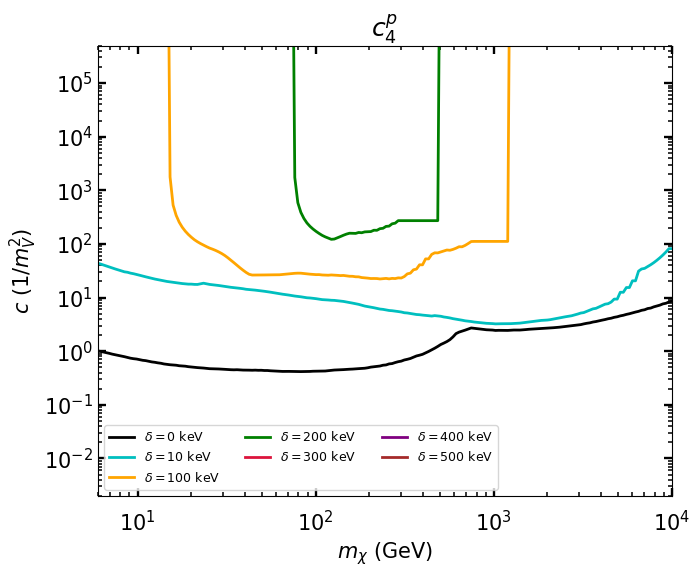}
\includegraphics[width=7.49cm,height=6cm]{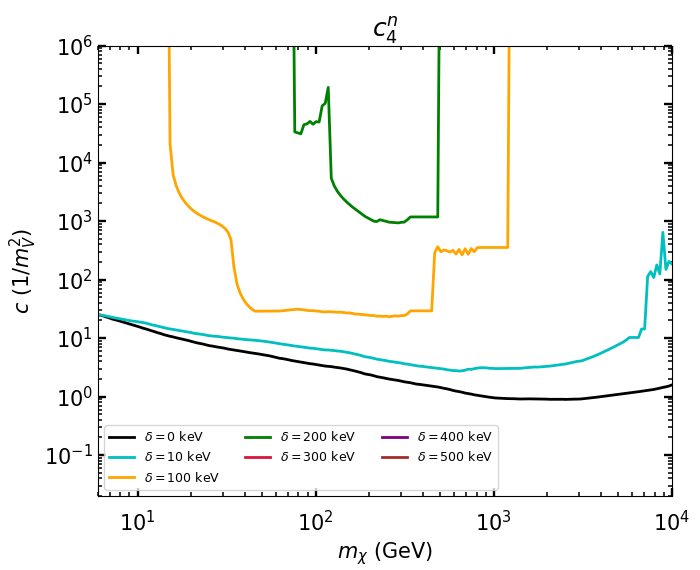}
\caption{Halo--independent upper limits on SI and SD interaction couplings as a function of $m_{\chi}$. 
Limits are shown for different choices of $\delta$.}
\label{fig:limit_mx_diff_delta}
\end{figure*}

In both plots $\delta_{\rm max}$ is determined by the intersection of the dark--green solid line for which 
$u^{\rm C-max}$ = $\min(u^{\rm DD-min},u_{\rm max})$ and the dashed blue line representing the maximal limit of $\delta$ from capture when $u$ =0. Specifically this latter value of $\delta$ is determined by the condition $E^\chi_{\rm max}\ge E^\chi_{\rm cap}$, an inequality that is always verified for $u$=0, provided that the inelastic scattering process is kinematically allowed. In particular this implies $w=v_{\rm esc}=v_{T*}$, with $v_{\rm esc}\equiv v_{\rm esc}(r=0)\simeq$ 1383 km/s the maximal escape velocity at the center of the Sun.  Moreover, in the SI case one can notice that $\delta_{\rm max}$ is above the light--green solid line representing the maximal reach of DD, $\delta$ = $\delta_{\rm max, DD}$. This implies that in the same plot $u^{\rm DD-min}>u_{\rm max}$, and on the dark--green solid line one has $u^{\rm C-max}$ = $u_{\rm max}$. Summarizing, for the SI case the same value of $\delta$ = $\delta_{\rm max}$ verifies at the same time the two conditions $w=v_{\rm esc}=v_{T*}$ (for $u=0$, implying 
$\delta_{\rm max}=1/2\mu_{\chi T}v_{\rm esc}^2$) and $u^{\rm C-max}$ = $u_{\rm max}$. In the $u$ -- $\delta$ plane the situation is schematically exemplified in Fig.~\ref{fig:u_delta_max}(a).
\noindent In particular, combining equation~(\ref{eq:E_max_min_condition}) for $u^{\rm C-max}$ with the condition $u^{\rm C-max}$ = $u_{\rm max}$ one gets:

\begin{eqnarray}
   && x^2\left [A^2(x,y)-1 \right ]=1\nonumber \\
   && A(x,y)\equiv \frac{1+y}{1-\frac{y}{\sqrt{1+x^2}}}\nonumber\\
   && x\equiv \frac{v_{\rm esc}}{u_{\rm max}}\simeq 1.76,\,\,\,\,\,y\equiv\frac{m_T}{m_\chi},
   \label{eq:y_value}
\end{eqnarray}

\noindent from which one obtains:

\begin{equation}
    y=\frac{\sqrt{1+x^2}-x}{1+x}\simeq 0.095,
\end{equation}

\noindent in a straightforward way. Notice that since $\delta_{\rm max,DD}<\delta_{\rm max}$ in this case $\delta_{\rm max}$ is only determined by capture in the Sun. On the other hand, for the SD case in the right--hand plot of Fig.~\ref{fig:delta_mx_scan} $\delta_{\rm max}<\delta_{\rm max,DD}$. The situation  is in this case depicted schematically in Fig.~\ref{fig:u_delta_max}(c), and, as a consequence, now $\delta_{\rm max}$ is determined by a combination of capture in the Sun and DD. Indicating with $N$ the nuclear target used in the DD experiment, for this to happen one needs $u^{\rm DD-min}\le u_{\rm max}$, with $u^{\rm DD-min}\ge v_{N*}$ (see Eq.~(\ref{eq:uth})). In particular this implies $v_{N *}\le u_{\rm max}$ and, as a consequence:

\begin{eqnarray}
&& \delta_{\rm max,DD}=\frac{1}{2}\mu_{\chi N} u_{\rm max}^2\ge\delta_{\rm max}, 
   \label{eq:vnstar1}
\end{eqnarray}

\noindent with $\mu_{\chi N}$ the $\chi$--$N$ reduced mass. Setting $r\equiv m_N/m_T$ and using Eq.~(\ref{eq:y_value}) the expression above yields:

\begin{eqnarray}
    && r = \frac{m_N}{m_T}\ge r_{\rm min}=\left(\frac{1+\sqrt{1+x^2}}{x^2(1+x)}-y(x)\right)^{-1}\simeq 3.9.
    \label{eq:mn_min}
\end{eqnarray}

\noindent In particular, when the condition above is {\it not} verified 
$\delta_{\rm max}$ is only determined by capture in the Sun, i.e. the DD bound is not strong enough to affect the maximal value of $\delta$, which is probed in a halo--independent way by NTs only. Specifically, in the SI case one has $T$ = $^{56}Fe$ and $N$ = $Xe$ (in LZ or XENON1T) with  $r=m_N/m_T\simeq$ 2.3 and indeed the condition~(\ref{eq:mn_min}) is not verified (notice that this is irrespective on whether the first condition in Eq.~(\ref{eq:uth}), that was used to obtain Eq.~(\ref{eq:mn_min}), holds)\footnote{When capture in the Sun is driven by $Fe$ not even the heaviest nuclear target used in DD, i.e. Tungsten in CRESST~\cite{cresst_2018}, can overcome the hierarchy of Eq.~(\ref{eq:mn_min}).}. On the other hand, for the SD interaction one has $T$ = $^{27}Al$ and either $N$ = $Xe$ (again in LZ or XENON1T) or $N$ = $I$ (in PICO--60($CF_3I$)) with $r=m_N/m_T \simeq$  4.86 or $r=m_N/m_T \simeq$  4.7. In this case the requirement~(\ref{eq:mn_min}) can be verified (if the experimental energy threshold is small enough, see again  Eq.~(\ref{eq:uth})). This is confirmed by the right--hand plot of Fig.~\ref{fig:delta_mx_scan}, where $\delta_{\rm max}$ is determined by a combination of DD and capture. We notice here that the fact that the sensitivity to inelastic scattering is enhanced for heavy nuclei has been known for a long time~(see for instance~\cite{halo_independent_Scopel_inelastic_2014} and references therein) so that the overall sensitivities to $\delta$ in Fig.~\ref{fig:delta_mx_scan} is improved using heavier targets for $T$ and/or $N$. On the other hand the lower bound of Eq.~(\ref{eq:mn_min}) represents a new result that pertains to the possibility for DD to compete with capture in the Sun to determine the maximal value of the mass splitting $\delta$ that can be probed in halo--independent way. In particular the inelastic WIMP--nucleus scattering process is kinematically favoured within the Sun compared to a DD experiment, thanks to the additional kick to the incoming velocity of the WIMP provided by the gravitational acceleration in the celestial body~\cite{wd_sogang_2023}. So the target mass hierarchy obtained in Eq.~(\ref{eq:mn_min}) quantifies how much heavier the nuclear target $N$ must be in a DD experiment in order to overcome the fact that in a terrestrial detector the incoming WIMPs are slower. Such required hierarchy depends on the astrophysical parameters $u_{\rm max}$ and $u_{\rm esc}$, while it does not directly depend on the details of the WIMP--nucleus interaction (although for a different type of interaction the targets $T$ and $N$ driving capture in the Sun and DD may change, leading to different kinematic consequences~\cite{halo_independent_sogang_2023}). In particular, for $u_{\rm max}>u_{\rm max}^{ref}$ the value of $r_{\rm min}$ in Eq.~(\ref{eq:mn_min}) is reduced (see Appendix~\ref{app:umax_dep} and Fig.~\ref{fig:umax_rmin}).  

The discussion above assumes $u_{\rm max}$ = $u_{\rm max}^{ref}$ = 780 km/s. In Appendix~\ref{app:umax_dep} we show that whenever for a given choice of $u_{\rm max}^{ref}$ the value of $\delta_{\rm max}$ is determined by a combination of capture and DD (as in the case of the right--hand plot of Fig.~\ref{fig:delta_mx_scan} for a SD interaction) taking $u_{\rm max}> u_{\rm max}^{ref}$ does not modify the kinematic regions where a halo--independent bound is possible. However, if instead for $u_{\rm max}$ = $u_{\rm max}^{ref}$ the value of $\delta_{\rm max}$ is only determined by capture (as in the case of the left--hand plot of Fig.~\ref{fig:delta_mx_scan} for a SI interaction) there exists a value $u_{\rm max}$ = $u_{\rm max}^{\rm DD-C}>u_{\rm max}^{ref}$ for which the situation depicted in the central plot of Fig.~\ref{fig:u_delta_max} is verified, i.e. where DD becomes sensitive enough to contribute to fixing $\delta_{\rm max}$. In this case, taking $u_{\rm max}^{ref}<u_{\rm max}<u_{\rm max}^{\rm DD-C}$ implies a reduction of the $m_\chi$ -- $\delta_{\rm max}$ parameter space to which a halo--independent analysis is sensitive. Such reduction is shown in the left--plot of Fig.~\ref{fig:delta_mx_scan} as a dashed area, where $u_{\rm max}^{\rm DD-C}\simeq$ 980 km/s. In the same plot the dashed light--green line represents $\delta_{\rm max, DD}$ for  $u_{\rm max}$ = $u_{\rm max}^{\rm DD-C}$. Taking $u_{\rm max}> u_{\rm max}^{\rm DD-C}$ does not change the red shaded region any further. On the other hand the kinematic regions shown in Fig.~\ref{fig:delta_mx_scan} are unchanged, and the other plots of this Section mildly modified, when the assumption of thermalization of the WIMP with the solar plasma is not made and the corresponding relaxation of the bounds is applied (see discussion in Appendix~\ref{app:thermalization}, where we show that the bounds are at most relaxed by about a factor of 6).

\begin{figure*}[ht!]
\centering
\includegraphics[width=7.49cm,height=6cm]{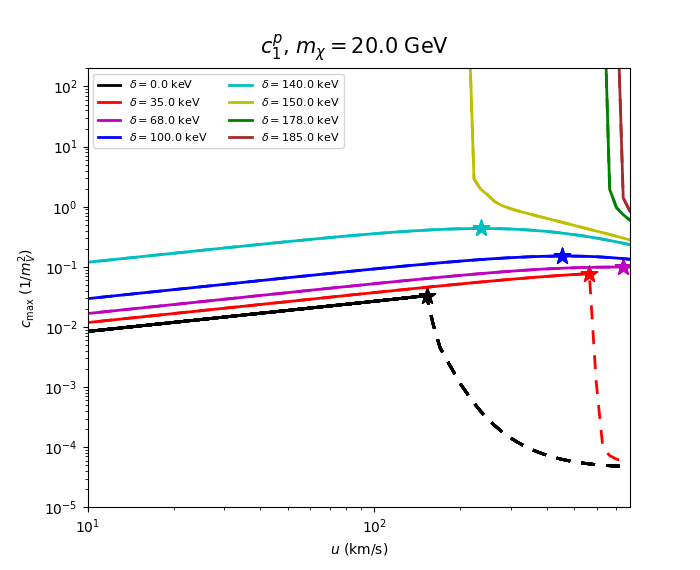}
\includegraphics[width=7.49cm,height=6cm]{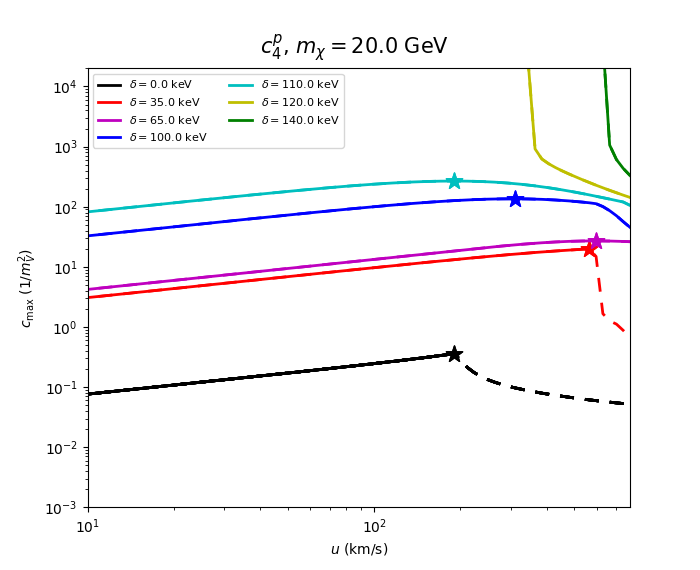}
\includegraphics[width=7.49cm,height=6cm]{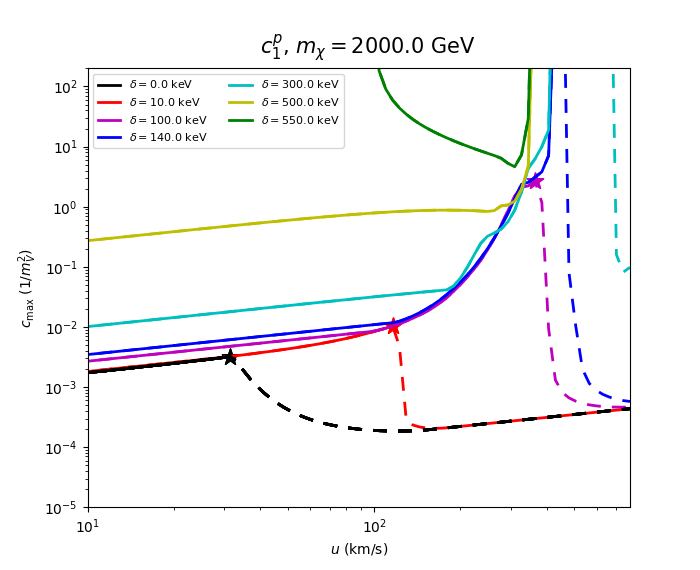}
\includegraphics[width=7.49cm,height=6cm]{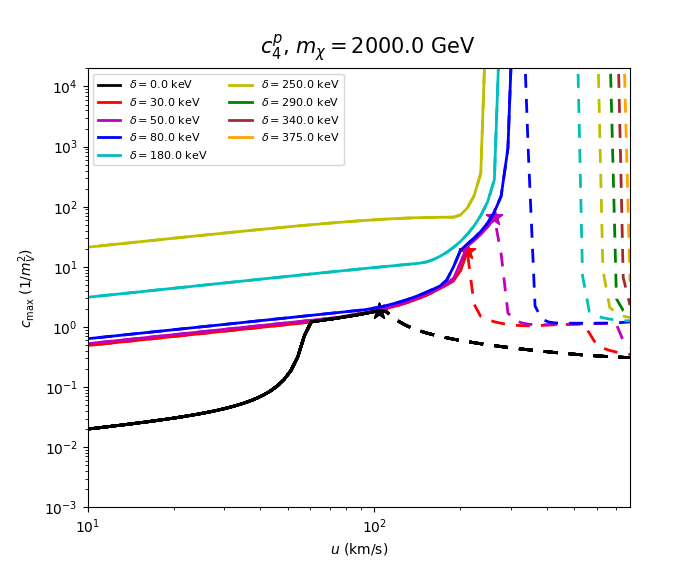}
\caption{For each curve (or the upper limit) the part dominated by the neutrino telescope observation is 
shown by the continuous line, while the part dominated by the direct detection is represented by the dashed line. The star on each curve indicates either the values of $\tilde{u}$ and $\tilde{c}$ defined in Eq.~\eqref{eq:u_tilde_c_tilde}, or the values of $u$ and $c$ defined in Eq.~(\ref{eq:limit_B}).}
\label{fig:ci_u_diff_delta}
\end{figure*}

\begin{figure}
\centering
\includegraphics[width=7.49cm,height=6cm]{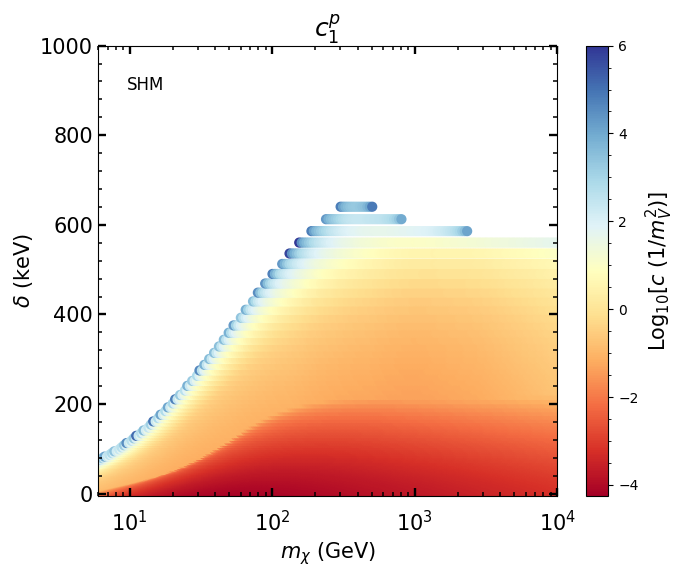}
\includegraphics[width=7.49cm,height=6cm]{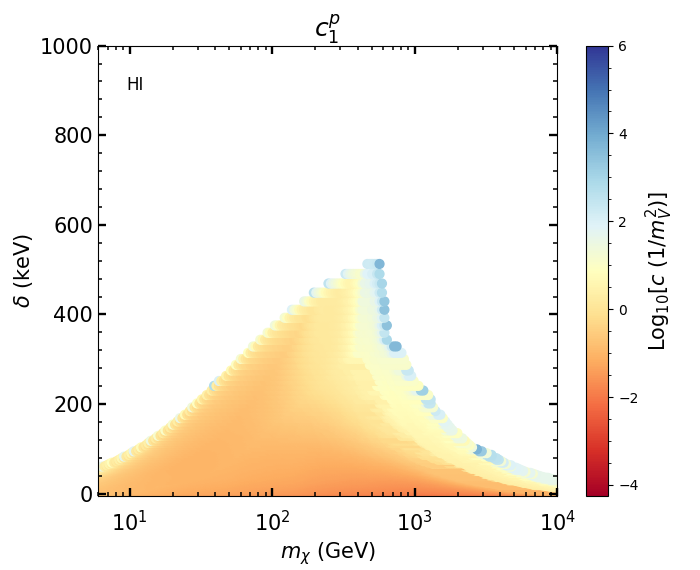}
\includegraphics[width=7.49cm,height=6cm]{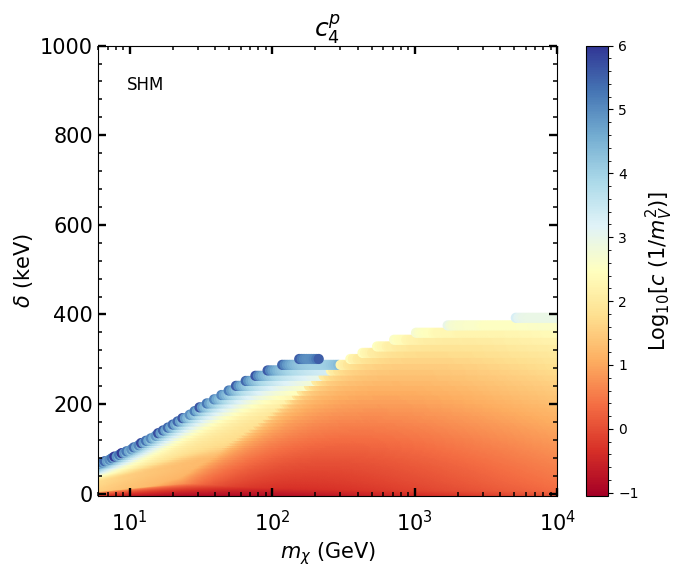}
\includegraphics[width=7.49cm,height=6cm]{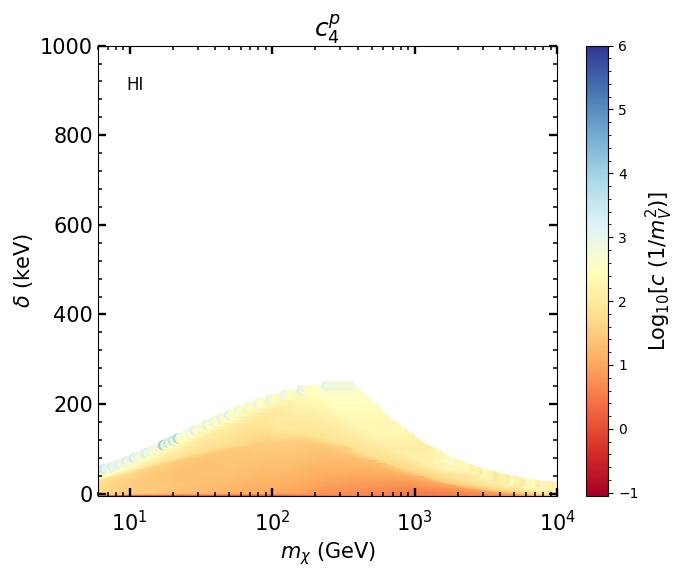}
\includegraphics[width=7.49cm,height=6cm]{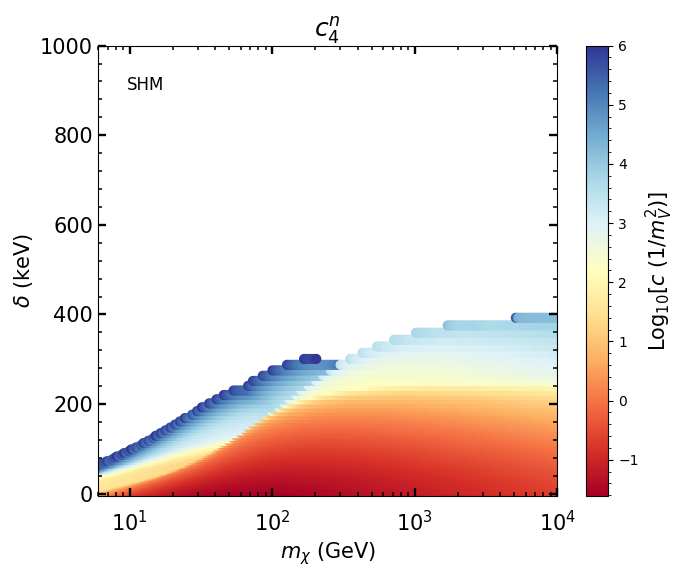}
\includegraphics[width=7.49cm,height=6cm]{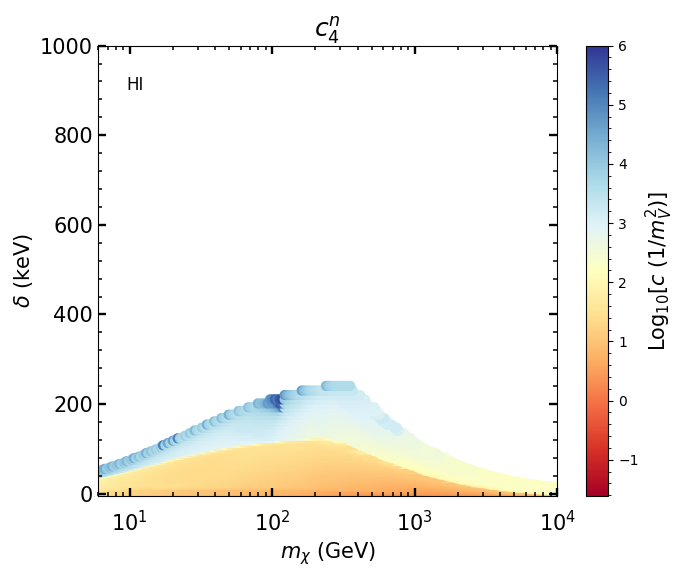}
\caption{Regions in the $m_\chi$ -- $\delta$ plane where a finite limit on the WIMP--nucleon 
coupling is obtained are shown by color shades for the SHM case (left column) and 
the halo--independent case (right column). The color code indicates the absolute values 
of the upper limit on the coupling (in $\rm Log_{10}$ scale) in the unit of $1/m^2_V$.
{\it Top}: considering the SI WIMP-proton coupling ($c^p_1$), 
{\it middle}: SD WIMP-proton coupling ($c^p_4$) and 
{\it bottom}: SD WIMP-neutron coupling ($c^n_4$).}
\label{fig:delta_mx_scan_absolute_c}
\end{figure}

To make contact to the usual way the bounds are presented, in Fig.~\ref{fig:limit_mx_diff_delta} we discuss the halo--independent exclusion plots of the SI and SD couplings as a function of the WIMP mass $m_\chi$. Since in such figure the bounds for $c_1^p$ and $c_1^n$ are indistinguishable only the case of $c_1^p$ is shown. In each plot the bounds are shown for different values of the mass splitting $\delta$. In particular, consistently with the previous discussion, for increasing values of $\delta$ the constraints get weaker and eventually the WIMP mass range for which a halo--independent bound is possible start shrinking, eventually converging to the value of $m_\chi$ that in Figs.~\ref{fig:delta_mx_scan} correspond to $\delta_{\rm max}$. In agreement with the kinematic discussion provided in Section~\ref{sec:combined_sensitivity_idm} such value of $m_\chi$ represents the boundary between a low--mass regime, where capture alone eventually determines the halo--independent upper bound on the coupling $c$, and a large mass one, where instead the same constraint is determined by both DD and capture in the Sun. This is explicitly shown in Fig.~\ref{fig:ci_u_diff_delta}, where for the same two values of the WIMP mass of Figs.~\ref{fig:delta_u_mchi_20_GeV} and \ref{fig:delta_u_mchi_2_TeV} (one in the low--mass regime, $m_\chi=20$ GeV and the other in the large mass regime, $m_\chi=2$ TeV) we plot as a function of $u$ and for different values of $\delta$ the quantities $(c^{\rm DD})_{\rm max}(u)$ and $(c^{\rm NT})_{\rm max}(u)$ that correspond to the bounds on the WIMP--proton couplings of Fig~\ref{fig:limit_mx_diff_delta}. In the same Figure the star represents 
either the values of $\tilde{u}$ and $\tilde{c}$ defined in Eq.~\eqref{eq:u_tilde_c_tilde}, or the values of $u$ and $c$ defined in Eq.~(\ref{eq:limit_B}).

Specifically, Fig.~\ref{fig:ci_u_diff_delta} shows how when $\delta$ =0 the quantity $\tilde{c}$ is obtained at low mass by exploiting the complementarity between DD and capture in the Sun, but when $\delta>$ 0 the sensitivity of DD is strongly reduced, so that only 
NT's determine the bound, as long as $\delta$ is not so large that inelastic scattering becomes kinematically not accessible  
even in the Sun, in spite of the much larger WIMP velocities. Notice that in this case $u^{\rm C-max}_T$ always exceeds $u_{\rm max}$, i.e. for $u<u_{\rm max}$ the WIMP is always captured if the inelastic scattering process takes place, and DD either plays a marginal role in determining the halo--independent bound or no role at all. On the other hand, at large WIMP masses Fig.~\ref{fig:ci_u_diff_delta} shows that now $u^{\rm C-max}_T< u_{\rm max}$, so that when the WIMPs are too fast they are not captured in the Sun: in this case DD is instrumental in probing higher WIMP velocities, as long as $u^{\rm DD-min}<u_{\rm max}$ and the inelastic scattering process is kinematically allowed in DD.  

We conclude by pointing out that the halo--independent bounds obtained with the single--stream method 
are probably overly pessimistic, since in order to remove the dependence on the velocity distribution one makes the somewhat extreme assumption that the full flux of the incoming WIMPs is concentrated at the speed corresponding to the smallest experimental sensitivity of the combination of DD experiments and Neutrino Telescopes. To give a quantitative assessment of this aspect,  in Fig.~\ref{fig:delta_mx_scan_absolute_c} we compare the contour plots in the $m_\chi$ -- $\delta$ plane of the upper limit on the couplings $c_1^p$, $c_4^p$ and $c_4^n$ (in units of $1/m^2_V$) obtained in the SHM (left column) to those obtained in the single--stream method (right column). Besides the kinematic effect already discussed in Fig.~\ref{fig:delta_mx_scan} (the SHM bounds extend to wider regions, especially at larger WIMP masses) one can observe that quantitatively the halo--independent bound is systematically much weaker than the SHM one (in the case of a SI interaction this effect is more pronounced in the range of moderate values of $\delta$ accessible to DD). The results of Fig.~\ref{fig:delta_mx_scan_absolute_c} are obtained for $u_{\rm max}$ = $u_{\rm max}^{ref}$. As shown in the left--hand plot of Fig.~\ref{fig:delta_mx_scan} in the case of a SI interaction taking  $u_{\rm max} > u_{\rm max}^{ref}$ reduces the kinematic region in the $m_\chi$ -- $\delta$ plane where a halo--independent single--stream bound is possible. As far as the quantitative bound on the coupling is concerned, in Ref.~\cite{halo_independent_sogang_2023}
it was shown that for the case of elastic scattering ($\delta$ = 0) increasing $u_{\rm max}$ by a factor of ten (i.e. to a value that largely exceeds the expectations for the escape speed in our Galaxy) weakens the single--stream bound by at most a factor of two, due to the suppression of the form factor in  the WIMP--nucleus scattering process  at large nuclear recoil energies. Since the form factor is a decreasing function of the recoil energy, and the latter is reduced when $\delta>$ 0 and the same value of the other parameters, we conclude that also for inelastic scattering when the single--stream bound is possible its quantitative weakening  for $u_{\rm max} > u_{\rm max}^{ref}$ is moderate, not exceeding a factor of two.

\section{Conclusions}
\label{sec:conclusions}

In the present paper we have made a quantitative assessment of the validity and applicability of the halo--independent single--stream method introduced in Ref.~\cite{Halo-independent_Ferrer2015} to the case of Inelastic Dark Matter. In particular we have shown that a non--vanishing mass splitting $\delta$ modifies the range of incoming velocities $u$ to which both DD and Capture in the Sun are sensitive, and that for particular combinations of $m_\chi$ and $\delta$ the complementarity of the two detection techniques required by the method is lost because when $\delta$ is too large they are not sensitive anymore to the full range of $u\le u_{\rm max}$. In such cases a halo--independent constraint on the WIMP--nucleus interaction is not possible, and a specific choice of the $f(u)$ is required to obtain a bound on the cross section or the coupling. Specifically, such values of $\delta$ depend on $m_\chi$ and are provided in Fig.~\ref{fig:delta_mx_scan} for both a SI and a SD interaction. Namely, assuming for the maximal speed $u_{\rm max}$ of the WIMP in the solar rest frame the reference value $u_{\rm max}^{ref}$ = 780 km/s, a halo--independent bound is possible when $\delta\lesssim$ 510 keV for a SI interaction\ and when $\delta\lesssim$ 245 keV for a SD interaction. We have also shown that in the case of a SI interaction the maximal value of $\delta$ that can be probed in a halo--independent way is slightly reduced to $\delta \lesssim$ 490 km/s when 780 km/s $\lesssim u_{\rm max}\lesssim$ 980 km/s. Moreover, when $u_{\rm max}>u_{\rm max}^{ref}$ the single--stream bounds can be weakened due to the suppression of the form factor in  the WIMP--nucleus scattering process  at large nuclear recoil energies. In particular the effect is at most a factor of two when $u_{\rm max}$ is increased by a factor of ten (i.e. $u_{\rm max}$ = 8000 km/s), a value that largely exceeds the expectations for the escape speed in our Galaxy.

On general grounds the inelasticity of the WIMP scattering process favours heavy targets. Moreover it disfavours DD compared to capture in the Sun, because in the Sun the WIMPs are much faster than in DD thanks to the gravitational acceleration in the celestial body. In particular in our analysis we have found that the combination of these two properties implies that DD can play a role in the determination of the maximal value of the mass splitting $\delta$ for which a halo--independent bound is possible only if $u_{\rm max}$ is large enough and/or if the target $m_N$ used in DD is heavier enough compared to the target $m_T$ that drives capture in the Sun, to compensate for the faster WIMPs inside the Sun. Specifically, for $u_{\rm max}$ = $u_{\rm max}^{ref}$ = 780 km/s such maximal value of $\delta$ is completely determined by capture in the Sun unless $m_N$ is larger than about four times $m_T$ (see Eq.~(\ref{eq:mn_min})). 

Two different scenarios arise at small and large WIMP mass $m_\chi$. In the low--mass regime 
the NT bound is always more constraining than that for DD and is sufficient alone to provide a halo--independent constraint. In this case a halo--independent bound is achievable as long as the WIMP--nucleus inelastic process is kinematically allowed. On the other hand at large WIMP masses the halo--independent bound is given by a combination of NT and DD. In this case an increasing value of the mass splitting $\delta$ favours again capture compared to DD because the reduction of the range of WIMP initial speeds $u$ probed by DD is more sensitive to $\delta$ than that for capture (compare Eqs.~(\ref{eq:e_max_dd}) and (\ref{eq:e_min_NT})). 

Quantitatively we find that the halo--independent bound can be much weaker than the SHM one. For the  SI interaction this effect is particularly pronounced in the range of moderate values of $\delta$ accessible to DD.

In our analysis we have assumed thermalization between WIMPs and the solar plasma. Numerical simulations~\cite{Blennow_2018} have put this into question, showing that the WIMPs may stop interacting with the nuclei in the Sun when their dispersion velocity is too low to trigger the inelastic process.
When the assumption of thermalization is not made the effect on our results is relatively mild (see Appendix~\ref{app:thermalization}).

\section*{Acknowledgements}
This research was supported by the National
Research Foundation of Korea (NRF) funded by the Ministry of Education
through the Center for Quantum Space Time (CQUeST) with grant number
2020R1A6A1A03047877 and by the Ministry of Science and ICT with grant
number 2021R1F1A1057119.

\appendix

\section{Implementations of experiments}
\label{app:experiments}

\subsection{LZ and XENON1T}
\label{app:lz}

LUX--ZEPLIN(LZ) has an exposure of 3.3$\times$10$^5$ kg days. We use the efficiency provided in Fig.~2 of~\cite{LZ_2022}.
To calculate our bounds we assume 3.4 residual candidate events in the nuclear recoil energy range 1.25 keV $\le E_R \le$ 80 keV~\cite{LZ_2022}, which reproduce the published exclusion plots for a standard SI interaction.
For XENON1T we assume 7 events in the nuclear recoil energy range 1.8 keV $\le E_R \le$ 62 keV~\cite{xenon_2018} and the efficiency provided in Fig.~1 of~\cite{xenon_2018} with an exposure of 3.6$\times$10$^{5}$ kg days.
The provided efficiencies are directly expressed in keV including the effects of quenching and energy resolution for both experiments.


\subsection{PICO--60 ($C_3F_8$)}
\label{app:pico60_c3f8}
PICO--60 is a bubble chamber that detects a signal only above some value $E_{\rm th}$ of the deposited energy.
In this case the expected number of events is given by:

\begin{equation}
R=N_T MT\int_0^{\infty} P(E_R) \frac{dR}{dE_R} dE_R,
\label{eq:r_threshold}
\end{equation}

\noindent with $P(E_R)$ the nucleation probability.

For the $C_3F_8$ target material we used the total exposure~\cite{pico60_2019}, consisting in 1404 kg day at the threshold $E_{\rm th}$=2.45 (with 3 observed candidate events and 1 event from the expected background, implying a 90\%C.L. upper bound of 6.42 events~\cite{feldman_cousin}) and 1167 kg day keV at the threshold $E_{\rm th}$=3.3 keV (with zero candidate events and negligible expected background, implying an upper bound of 2.3 events at 90\% C.L.).
We have assumed the nucleation probabilities in Fig. 3 of \cite{pico60_2019} for the two runs. 

\subsection{PICO--60 ($CF_3I$)}
\label{app:pico60cf3i}

For the PICO--60 run employing a $CF_3I$ target we adopt an energy threshold of 13.6 keV and a 1335 kg days exposure. The nucleation probabilities for each target element are taken from Fig.4 in~\cite{pico60_2015}.

\subsection{Neutrino Telescopes}
\label{exp:NT}
Neutrino telescopes provide constraints on the neutrino flux from the 
annihilation of WIMPs captured in the Sun. In this work we have used the results of the observations of 
the neutrino telescopes IceCube \cite{IceCube:2016, IceCube:2021_LE} and Super--Kamiokande \cite{SuperK_2015}. 
Analysing the neutrino data taken from the direction of the Sun for a lifetime of 532 days 
the IceCube collaboration has provided 90\% C.L. upper bounds on the WIMP annihilation rate 
$\Gamma_\odot$ for different annihilation channels ($b\bar{b}$, $W^+W^-$ and $\tau^+\tau^-$) \cite{IceCube:2016}. 
For example, considering the $b\bar{b}$ channel, the IceCube bound is 
$\Gamma_\odot\lesssim [7.4\times10^{24} \rm s^{-1}, 7.3\times10^{20} \rm s^{-1}]$ 
for $m_{\chi}$ in the range 35 GeV -- 10 TeV. 
The bounds from the Super--Kamiokande collaboration~\cite{SuperK_2015} are obtained 
using the data for an exposure of 3903 days. Such Super--Kamiokande bounds, 
which are expressed in terms of 95\% C.L. upper limits on the WIMP--nucleon cross section 
for $m_{\chi}$ in the range 6 -- 200 GeV, correspond to $\Gamma_\odot \lesssim [1.2\times10^{25} \rm s^{-1}, 1.2\times10^{23} \rm s^{-1}] $ for the $b\bar{b}$ annihilation channel. 
In our work, with the goal to obtain conservative bounds, we consider only 
annihilations to the $b\bar{b}$ channel, which, among the different channels usually studied by neutrino telescopes, provides the smallest neutrino flux\footnote{Annihilation to light quarks, which are stopped in the solar plasma before hadronizing and decaying, have also been considered in the literature, and are particularly challenging to NT's~\cite{capture_light_quarks1,capture_light_quarks2,capture_light_quarks3, NT_DD_Blennow2015}.}.


\section{Thermalization of IDM within the Sun}
\label{app:thermalization}

\begin{figure*}[ht!]
\centering
\includegraphics[width=7.49cm,height=6cm]{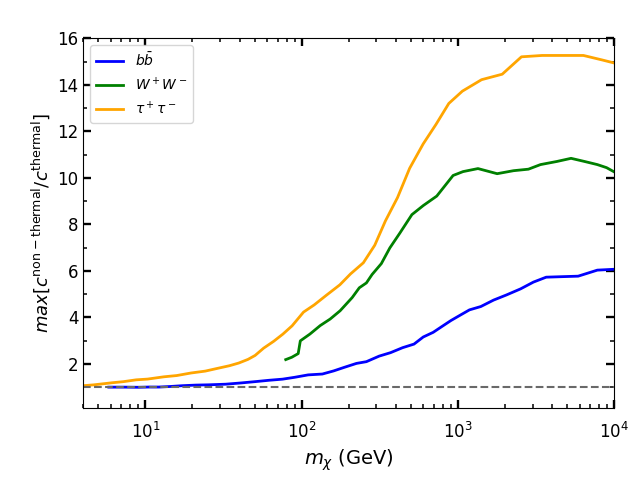} \\
\caption{Maximal relaxation of the NT bounds when the assumption of thermalization of the WIMPs with the solar plasma is relaxed, and Eq.~(\ref{eq:ca_non_thermal}) is used instead of~(\ref{eq:ca_thermalization}) for the determination of the relaxation time between capture and annihilation. The three annihilation channels $b\bar{b}$,  $W^+W^-$ and $\tau^+\tau^-$ are shown. The relaxation of the halo--independent bound can be smaller when the NT bound is combined with DD.}
\label{fig:relax_nonthermal}
\end{figure*}

Assuming that the WIMPs thermalize with the solar plasma and $\langle \sigma v\rangle$ = $3\times 10^{-26}$ cm$^3$ s$^{-1}$, the present bounds on the annihilation rate $\Gamma_\odot$ from NTs imply $\tau_\odot\ll t_\odot$ and $\Gamma_\odot$ = $C_\odot/2$ (equilibrium between capture and annihilation, a quantitative discussion on this is provided in Appendix (A.4) of \cite{halo_independent_sogang_2023}).  
This allows to use the results of neutrino searches to directly put constraints on the capture rate $C_\odot$ and hence on the WIMP--nucleon couplings that drive the WIMP capture. However, in the case of inelastic scattering numerical simulations~\cite{Blennow_2018} show that the WIMPs may stop interacting with the nuclei in the Sun when their dispersion velocity is too low to trigger the inelastic process. In this case it has been shown that for a Maxwellian distribution $f(u)$ the WIMPs thermalize, but at a temperature $T_\chi>T_c$, leading to a shallower density profile inside the Sun, suppressing the annihilation rate $\Gamma_\odot$ when equilibrium with capture is not reached~\cite{Blennow_2018}. To take the effect of thermalization into account in a halo--independent way appears challenging, requiring in principle to perform numerical simulations for a large set of initial velocity distributions and in the full $m_\chi$ -- $\delta$ parameter space. However one can notice that the most conservative bounds are obtained by minimizing the $C_A$ quantity defined in Eq.~(\ref{eq:ca}), which fully contains the effect of thermalization. $C_A$ does not depend on the number of captured WIMPs (which is fixed by the capture rate) but only on the density profile of the WIMPs inside the Sun. In particular, for a fixed number of WIMP inside the Sun the worst--case scenario leading to the minimal annihilation signal corresponds to a constant density profile, $n_\chi(r)$ = $n_0$, for which:

\begin{equation}
  C_A = \frac{\langle \sigma v \rangle}{V_\odot},
  \label{eq:ca_non_thermal}
\end{equation}

\noindent with $V_\odot$ the volume of the Sun. Numerically, this implies that $C_A$ drops between 3 and 8 orders of magnitude below the values obtained using Eq.~(\ref{eq:ca_thermalization}), when thermal equilibrium between the WIMPs and the solar plasma is assumed, leading to values of the equilibration time $\tau_\odot$ between 2 and 4 orders of magnitude longer (see Eq.~(\ref{eq:t_eq})). In this case equilibrium between capture and annihilation is not reached anymore, and indicating with $c^{thermal}$ the NT bound on the coupling obtained using for $C_A$ Eq.~(\ref{eq:ca_thermalization}), and with $c^{non-thermal}$ the same quantity obtained using for $C_A$ Eq.~(\ref{eq:ca_non_thermal}), one gets:

\begin{equation}
    \frac{c^{non-thermal}}{c^{thermal}}\simeq \left (\frac{V_\odot}{2 \langle \sigma v\rangle t_\odot^2\Gamma_{\rm exp} }  \right )^{1/4},
    \label{eq:relax_nonthermal}
\end{equation}

\noindent with $\Gamma_{\rm exp}$ the experimental upper bound on the WIMP annihilation rate, and approximating in Eq.~(\ref{eq:gamma_rate}) the hyperbolic tangent with its argument. As shown in Fig.~\ref{fig:relax_nonthermal} this leads to a relaxation of the NT bound up to a factor of about 6 in the case of the $b\bar{b}$ annihilation channel, and somewhat higher for $W^+W^-$ and $\tau^+\tau^-$. Such relatively mild effect can be ascribed both to the level of the present experimental bounds and to the 1/4 power in Eq.~(\ref{eq:relax_nonthermal}). The relaxation of the halo--independent bound can be smaller when the NT bound is combined to DD, following the procedure outlined in Section~\ref{sec:single_stream}.

\section{Dependence on $u_{\rm max}$}
\label{app:umax_dep}

In the discussion of Section~\ref{sec:analysis} we adopted the standard value $u_{\rm max}$ = $u_{\rm max}^{ref}$ = 780 km/s. Here we wish to discuss how such considerations can be modified by adopting $u_{\rm max}>u_{\rm max}^{ref}$.

\begin{figure*}[ht!]
\centering
\includegraphics[width=7.49cm,height=6cm]{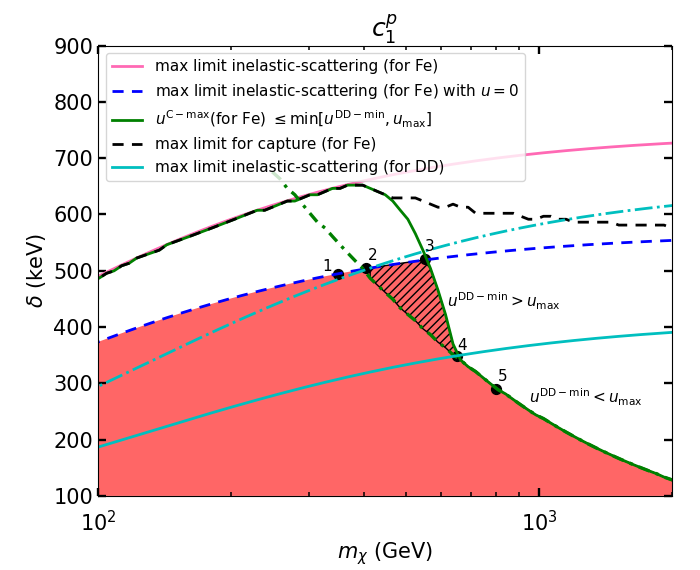} \\
\caption{Zoom--in of the $m_\chi$ -- $\delta$ parameter space shown in the left--hand plot of Fig.~\ref{fig:delta_mx_scan} for a SI interaction. The numbers indicate the benchmarks whose corresponding $u$ -- $\delta$ parameter space is shown in each of the five plots of Fig.~\ref{fig:delta_mx_zoom_benchmarks}.}
\label{fig:delta_mx_zoom}
\end{figure*}

\begin{figure*}[ht!]
\includegraphics[width=16cm,height=3.5cm]{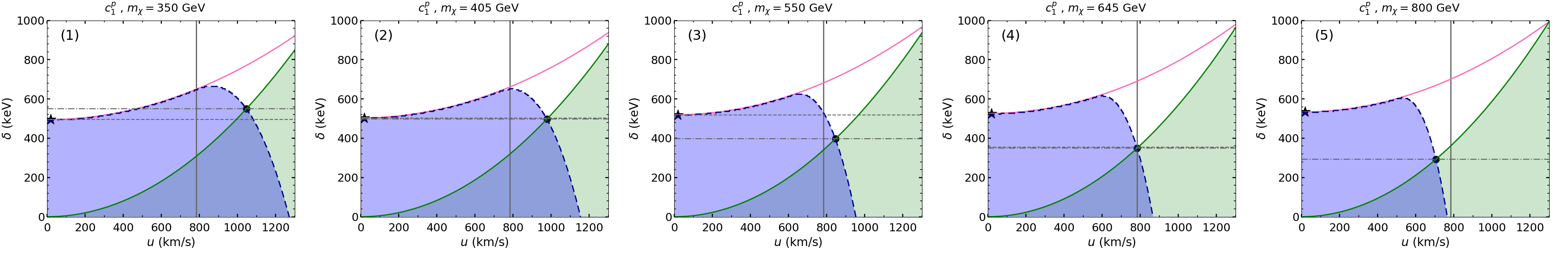}
\caption{$u$ -- $\delta$ parameter space for each of the five benchmarks indicated with numbers if Fig.~\ref{fig:delta_mx_zoom}.}
\label{fig:delta_mx_zoom_benchmarks}
\end{figure*}

In order to do so in Fig.~\ref{fig:delta_mx_zoom} we provide a zoom--in of the $m_\chi$ -- $\delta$ parameter space shown in the left--hand plot of Fig.~\ref{fig:delta_mx_scan} for a SI interaction. For $u_{\rm max}$ = $u_{\rm max}^{ref}$ in such plot the parameter space for which a single--stream halo--independent bound on the $c_1^p$ coupling is possible is shaded in red and is delimited from above by the green solid line where $u^{\rm C-max}$ = $\min(u^{\rm DD-min},u_{\rm max})$, and by the dashed blue line that represents the maximal $\delta$ for which inelastic scattering on $Fe$ is possible when $u=0$. In the same plot the numbers indicate the benchmarks whose corresponding $u$ - $\delta$ parameter space is shown in each of the five plots of Fig.~\ref{fig:delta_mx_zoom_benchmarks}.  
In such plots, and in analogy to Figs.~\ref{fig:delta_u_mchi_20_GeV} and \ref{fig:delta_u_mchi_2_TeV}, at fixed $m_\chi$ the blue--shaded region represents the parameter space kinematically accessible to capture and the green--shaded one that accessible to DD. Moreover, the vertical solid line represents the reference value $u_{\rm max}^{ref}$ for $u_{\rm max}$. 

In particular, in all the plots of Fig.~\ref{fig:delta_mx_zoom_benchmarks} the filled dot represents the values $\hat{u}$, $\hat{\delta}$ at the intersection between $\delta_{\rm max, C}$ (the upper boundaries of the blue--shaded region) and $\delta_{\rm max, DD}$ (the upper boundary of the green--shaded one) while the star represents $\delta_{u=0}$, the maximal value of $\delta$ for which capture is possible when $u$ = 0. In all the plots a halo--independent bound is possible for $\delta<\min(\hat{\delta},\delta_{u=0})$. Such upper bound is indicated with a horizontal dotted line if the additional constraint $u<u_{\rm max}^{ref}$ is imposed, while when $u$ is allowed to extend beyond $u_{\rm max}^{ref}$ the bound is given by the minimum between the horizontal dashed line and the horizontal dotted line. 

\begin{figure*}[ht!]
\centering
\includegraphics[width=7.49cm]{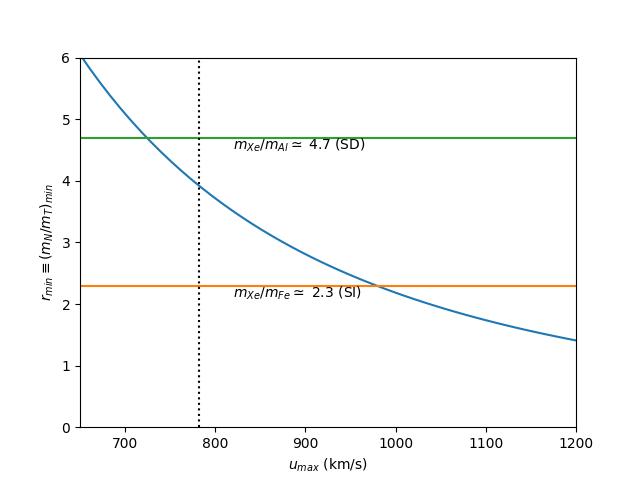}
\caption{Expression of $r_{\min}\equiv \left(m_N/m_T\right )_{\rm min}$ (with $N$ the target driving DD and $T$ the target driving capture in the Sun) of Eq.(\ref{eq:mn_min}) as a function of $u_{\rm max}$. The upper horizontal line represents $r=m_{Xe}/m_{Al}\simeq$ 4.86 for the SD interaction, and the lower horizontal line is   $r=m_{Xe}/m_{Fe}\simeq$ 2.3 for the SI interaction. The intersections between $r_{\min}$ and the two horizontal lines determine the value of the speed $u_{\rm max}^{\rm DD-C}$ defined in Appendix~\ref{app:umax_dep} for the SD and for the SI interaction, respectively.}
\label{fig:umax_rmin}
\end{figure*}

Let's consider in detail the five benchmarks:

\begin{itemize}
\item{\bf Benchmark (1):} when the WIMP mass is low enough the sensitivity of DD is reduced, and $\delta_{u=0}<\hat{\delta}$, with $\hat{u}>u_{\rm max}^{ref}$. In this case the upper bound on $\delta$ is only determined by capture and equal to $\delta_{u=0}$ for any value of $u_{\rm max}$. {\it In particular such boundary does not change when $u_{\rm max}>u_{\rm max}^{ref}$}. 

\item{\bf Benchmark (2):} There always exists a specific value of $m_{\chi}$ for which $\hat{\delta}$ = $\delta_{u=0}$. In such case 
$\delta<\delta_{u=0}=\hat{\delta}$, and DD starts becoming sensitive enough to contribute to the constraint on $\delta$. For the specific case of a SI interaction this happens when $\hat{u}\equiv u_{\rm max}^{\rm DD-C}\simeq$ 980 km/s, while for a SD interaction $u_{\rm max}^{\rm DD-C}\simeq$ 724 km/s (see Fig.~\ref{fig:umax_rmin}).

\item{\bf Benchmark (3):} For higher values of $m_\chi$ one has $\hat{\delta}<\delta_{u=0}$ and the upper bound on $\delta$ is given by a combination of DD and Capture. In this case, if $u_{\rm max}^{\rm DD-C}>u_{\rm max}^{ref}$ (as for the SI interaction discussed in Fig.~\ref{fig:delta_mx_zoom_benchmarks}) one has $\delta<\delta_{u=0}$ (horizontal dotted line) if $u_{\rm max}$ = $u_{\rm max}^{ref}$, while $\delta<\hat{\delta}$ (horizontal dashed line) for $u_{\rm max}>u_{\rm max}^{ref}$. As a consequence {\it the upper value of $\delta$ keeps reducing when $u_{\rm max}$ is increased between $u_{\rm max}^{ref}$ and $\hat{u}<u_{\rm max}^{\rm DD-C}$}.  

\item{\bf Benchmark (4):} Further increasing $m_\chi$ eventually one has $\hat{u}$ = $u_{\rm max}^{ref}$.

\item{\bf Benchmark (5):} At even higher WIMP masses one has $\delta<\hat{\delta}$ for $\hat{u}<u_{\rm max}^{ref}$. In this case 
{\it taking $u_{\rm max}>u_{\rm max}^{ref}$ does not affect the bound}, which is equal to $\hat{\delta}$ for any $u$. This is also the only possible situation when $u_{\rm max}^{\rm DD-C}<u_{\rm max}^{ref}$, as in the specific case of the SD interaction shown in the right--hand plot of Fig.~\ref{fig:delta_mx_scan}. For this reason the SD case is not affected by taking  $u_{\rm max}>u_{\rm max}^{ref}$.
\end{itemize}

Summarizing the discussion above, the bound on $\delta$ is modified by taking $u_{\rm max}>u_{\rm max}^{ref}$ only when $u_{\rm max}^{ref}<u_{\rm max}^{\rm DD-C}$ and the upper bound on $\delta$ is provided by capture alone for $u<u_{\rm max}$ = $u_{\max}^{ref}$, a situation that, if present, only affects a limited range of $m_\chi$, where the maximal values of $\delta$ that can be probed in a halo--independent way are reached. One can easily determine if this happens for a given type of interaction by plotting the quantity $r_{\rm min}$ of Eq.~(\ref{eq:mn_min}) as a function of $u_{\rm max}$. This is done in Fig.~\ref{fig:umax_rmin}, where the intersection with the upper horizontal line for $r=m_{Xe}/m_{Al}\simeq$ 4.86 determines for the SD interaction $u_{\rm max}^{\rm DD-C}\simeq$ 724 km/s $<u_{\rm max}^{ref}$ (no dependence on  $u_{\rm max}>u_{\rm max}^{ref}$), while the intersection  with the lower horizontal line for  $r=m_{Xe}/m_{Fe}\simeq$ 2.3 determines for the SI interaction $u_{\rm max}^{\rm DD-C}\simeq$ 978 km/s $>u_{\rm max}^{ref}$ (dependence on $u_{\rm max}>u_{\rm max}^{ref}$). This straightforward procedure allows in principle to determine if the maximal value of $\delta$ is affected by $u_{\rm max} $ for any interaction besides SI or SD.

\end{document}